%% file: bir-imhd-v2.tex
\documentclass[a4paper,12pt]{article}
\pdfoutput=1
\usepackage{jheppub}
\usepackage[T1]{fontenc} 
\usepackage[utf8]{inputenc}
\newcommand{\mathsym}[1]{{}}

\usepackage{xspace}
\usepackage{float}

\usepackage{verbatim}
\usepackage{amsmath}
\usepackage{amssymb}
\usepackage{url}
\usepackage{bm} 
\usepackage{etoolbox}
\usepackage{xspace}
\usepackage{xpatch}
\xpatchcmd\section{\large}{\Large}{}{} 
\usepackage{epsfig,multicol,bbm}
\usepackage{stackrel}
\usepackage[normalem]{ulem}
\usepackage{hepunits}
\include{hep-commands}
\title{
	Generalization of Bantilan-Ishi-Romatschke flow to Magnetohydrodynamics
}
\author[a]{M. Shokri}
\affiliation[a]{IPM, School of Particles and Accelerators,
P.O. Box 19395-5531, Tehran, Iran}
\emailAdd{mshokri@ipm.ir}
\date{\today}
\abstract{
	We present a generalization of the Bantilan-Ishi-Romatschke (BIR) solution of relativistic hydrodynamics to relativistic magnetohydrodynamics (RMHD). Using the symmetries of the boundary of the Kerr-AdS5 black hole, and certain simplifying assumptions we solve the equations of RMHD on this boundary for a highly conductive fluid. We then transform the resulting solution to the flat spacetime. Furthermore, we show that the force-free condition causes the magnetic field to become singular at particular points and propose a regularization process for removing the singularities. The regularization process reveals the importance of non-vanishing electrical current in RMHD.
}
\arxivnumber{}
\begin{document}
	\maketitle
	\flushbottom
	\section{Introduction}\label{sec:intro}
	\par
	It is expected that high energy collisions of heavy ions produce enormous electromagnetic fields, probably stronger than any other known electromagnetic fields in nature~\cite{Rafelski:1975rf,Kharzeev-CME}. Simulations suggest that the produced magnetic field is on order of $10^{18}\!-\!10^{20}$ Gauss and linearly proportional to the collision energy~\cite{skokov2009,Voronyuk:2011jd,Bzdak:2011yy,Deng:2012pc}. The estimated magnetic characteristic length~\cite{Fukushima:2018grm}, i.e. $1/\sqrt{eB}$\footnote{Here $B$ is the magnitude of the magnetic field and $e$ is proton's electric charge.}, at its peak is comparable to the typical length scale of strong interactions at low energies, namely $1/\Lambda_{\text{QCD}}$. This remark has led to dozens of studies on the probable effects of the magnetic field on the observables of the heavy-ion collisions~\cite{Hattori:2016emy,Ghosh:2018xhh,Mukherjee:2017dls,Sadooghi:2016jyf,Fukushima:2015wck}. Most of the aforementioned works are based on the assumption of a static long-lasting magnetic field. This assumption, however, may not be realistic~\cite{Huang-review} and we can understand its unrealistic essence by a simple estimation. Consider $eB$ being of the order of $m_\pi^2$\footnote{$m_\pi$ is Pion's mass.}  in the timescales that the QGP is formed. Such an assumption gives rise to an Alfven velocity~\cite{gedalin1993} of the same order as the speed of sound. If that would be the case, the electromagnetic forces could compete with the pressure gradient in the determination of fluid kinematics. Since relativistic hydrodynamics is in a good agreement with the experimental data, one can infer that this is not the case and the Alfven speed must be much smaller than the speed of sound. This rough argument is confirmed by the results of different simulations of the early time dynamics of the electromagnetic fields that agree on the short lifetime of magnetic fields~\cite{skokov2009, Voronyuk:2011jd, Bzdak:2011yy, Deng:2012pc}. In addition to such effects, the magnetic field is also proposed as a probe to the chiral properties of the QGP~\cite{Kharzeev-CME}. It is suggested that the produced QGP may possess a chiral imbalance on an event-by-event basis that gives rise to a non-dissipative electric current in the direction of the magnetic field and consequently the separation of opposite charges. The charge separation produces an electric dipole which induces a difference between the distribution of final particles of opposite charges. This phenomenon is called the chiral magnetic effect (CME). The search for the CME is still a work in progress~\cite{Kharzeev-Status,Zhao:2018ixy}.
	\par
 At early times the medium is dominated by the gluons and almost a perfect insulator~\cite{Huang-review}. In such an insulator the magnetic field decays quickly. However, as the QGP forms and hydrodynamization happens, the mutual evolution of the QGP and electromagnetic fields is effectively understood using relativistic magnetohydrodynamics (RMHD). It has been suggested that since QGP is highly conductive it may significantly slow down the magnetic fields decay~\cite{Tuchin:2013ie}. From the RMHD perspective, this is understood by assuming the ideal MHD (iMHD) limit. In such a limit the electric field is suppressed and the magnetic field is decayed through the separation of field lines due to fluid's expansion~\cite{beckenstein1978}. The generalization of Bjorken flow to iMHD confirms that the lifetime of the magnetic field significantly enhances in such a regime~\cite{RPRR-Bjorken-iMHD, RPRR-Bjorken-iMHD-Magnetized}. However, as argued in~\cite{rotating-trans-mhd} the electric field in the Bjorken setup is not suppressed unless the electrical conductivity is much larger than the corresponding proper time, i.e. $\sigma_e\tau\gg1$. Such a condition is not satisfied by Lattice QCD (LQCD) results~\cite{LQCD-Sigma} and one may deduce that the iMHD is a poor approximation for the QGP. Furthermore, it is argued that the initial configuration of the fields, i.e. rotation in the terminology of~\cite{rotating-trans-mhd,cs-trans-mhd}, is dominant over the transport coefficients of the QGP~\cite{rotating-trans-mhd}. The results of such a simplified analytical model, which may be crucial in the search for CME, need to be justified by numerical investigations. Despite significant progress in  iMHD~\cite{Inghirami:2016iru,Inghirami:2018ziv,Inghirami:2019mkc}, numerical computation of resistive RMHD in the QGP context is still missing. 
	\par
	Analytical solutions of RMHD may provide insights into the evolution of electromagnetic fields in the QGP. As mentioned earlier the generalization of Bjorken flow to iMHD appeared in~\cite{RPRR-Bjorken-iMHD}. In~\cite{rotating-trans-mhd}, the assumption of infinite conductivity is relaxed and in~\cite{cs-trans-mhd} the results of the aforementioned work are generalized to the chiral magnetohydrodynamics. The method of non-conserved charges which was introduced in~\cite{rotating-trans-mhd}, is also utilized in~\cite{rotating-trans-mhd} for investigation of the CME from an RMHD perspective. The generalization of 3+1 self-similar~\cite{CGHK-cyl-flow} and Gubser~\cite{Gubser-Conformal,Gubser-Symmetry} flows to iMHD is studied in~\cite{cmhd} using a systematic procedure based on symmetry arguments. In all of these works, a force-free condition is assumed. Namely, it is assumed that the electromagnetic forces do not disturb the kinematic of the fluid. This can be regarded as a first approximation if the Aflven velocity is much smaller than the sound velocity. As it is discussed earlier, such an assumption is in agreement with the successes of relativistic hydrodynamics. A different approach is assumed in~\cite{Accelerated-Pu,Accelerated-She}, in which the kinematical modifications of the fluid due to external magnetic fields are studied. In these works, the magnetic fields are assumed to be external in the sense that they need not satisfy the Maxwell equations,
	Recently a novel exact analytical solution to relativistic hydrodynamics is presented in~\cite{bir} that breaks rotational and longitudinal boost symmetries that are assumed by Bjorken and Gubser flows. The aforementioned solution, which will be abbreviated as BIR\footnote{Bantilan-Ishi-Romatschke} flow from now on, is based on the duality of a rotating Kerr-Ads5 black hole with a peripheral heavy-ion collision in the four-dimensional flat spacetime. The equations of fluid mechanics are solved on the $S^3\times\mathbb{R}$ boundary of the black hole using symmetry arguments, and then Weyl transformed into the flat spacetime to obtain the solution. We present a generalization of BIR flow to iMHD in the current paper.
	\par
	The organization of this paper is as follows: In Sec.~\ref{sec:bir-hydro}, we briefly review the BIR solution. In Sec.~\ref{sec:s3r-mhd}, we solve the equations of RMHD on the $S^3\times\mathbb{R}$ boundary. After applying the symmetry arguments the iMHD limit is considered. The resulting equations are fewer than the number of unknowns. Hence, we make two additional assumptions: The force-free condition and local electrical neutrality, which help us find the field tensor and consequently the magnetic field. The section is closed by calculation of the gauge potential and magnetic helicity. In Sec.~\ref{sec:flat-imhd}, we transform the magnetic field into the flat spacetime and discuss some aspects of its behavior. Furthermore, we transform the magnetic helicity to the flat spacetime. In Sec.~\ref{sec:regulaization}, we briefly discuss singularities and provide an approximation to resolve them and in Sec.~\ref{sec:results}, we depict some of the results. The paper is concluded in Sec.~\ref{sec:conclusion}.
	\par
	In this paper, the natural units are used in which $\hbar=c=1$. The metric signature convention is mostly plus. The notations are adapted from~\cite{cmhd}, and the reader may consult Appendix A of the aforementioned work for more details. The differential forms are denoted by non-italic bold symbols such as $\dform{A}$, and $\dext$ is the exterior derivative. In utilizing the language of the differential forms we mainly follow~\cite{Markakis:2016udr}.
	\section{BIR solution to hydrodynamics}\label{sec:bir-hydro}
	\setcounter{equation}{0}
	\par
	In this section, we briefly review the BIR solution. This solution is obtained from an assumed duality between a particular solution to Einstein equations in five-dimensional spacetime, namely the Kerr-AdS5 solution, with the solution to the equations of fluid mechanics on the four-dimensional boundary. As it is shown in~\cite{bir}, the line element of Kerr-AdS5 has the following leading behavior at $r'\to\infty$ 
	\begin{equation*}
		\frac{ds_5^2}{L^2} \simeq -(1+r'^2) dt'^2 + \frac{dr'^2}{1+r'^2} + r'^2 \left(d\theta'^2 + \sin^2 \theta' \, d\phi'^2 + \cos^2 \theta' d\chi'^2 \right) + \orderof{(1/r'^2)}.
	\end{equation*}
	Consequently, the line element at the boundary of the spacetime is
	\begin{equation}\label{eq:s3xr-metric}
		\rd{s^2}=L^2\left[-\rd{t'^2}+\rd{\theta'^2}+\sin^2\theta'\rd{\phi'^2}+
		\cos^2\theta'\rd{\chi'^2}\right]\,,\qquad x^a = (t',\theta',\phi',\chi').
	\end{equation}
	The \eqref{eq:s3xr-metric} is the line element of a $S^3\times \mathbb{R}$ spacetime. To recognize this, let us use the following coordinate system:
	\begin{equation}\label{eq:s3r-cartesin}
		X=L\sin\theta'\cos\phi'\,,\quad Y=L\sin\theta'\sin\phi'\,,\quad Z=L\cos\theta'\cos\chi'\,,\quad V=L\cos\theta'\sin\chi'.
	\end{equation}
	The above coordinates satisfy the $S^3$ equation, namely
	\begin{equation}\label{eq:s3-eqn}
		X^2+Y^2+Z^2+V^2=L^2.
	\end{equation} 
	The conformal fluid mechanics of such a spacetime is studied in~\cite{S3xR-Minwalla} and is reviewed here. The fluid energy-momentum tensor is given by~\cite{Rezzolla}
	\begin{equation}\label{eq:fem-tensor}
		T^{ab}=\epsilon u^a u^b + p \Delta^{ab}+T^{ab}_{\text{dissipative}}.
	\end{equation}
	Here $u^a$ is the fluid's four-velocity, $p$ and $\epsilon$ are fluid's pressure and energy density respectively and $\Delta_{ab}\equiv g_{ab}+u_a u_b$ is the projection tensor. At first order~\cite{S3xR-Minwalla,Rezzolla}, the dissipative part is
	\begin{equation}\label{eq:first-order}
		T^{ab}_{\text{dissipative}}=-\zeta \left(\nabla_cu^c\right) \Delta^{ab}-2\eta\sigma^{ab}+q^a u^b+q^b u^a.
	\end{equation}
	Conformal invariance requires that the bulk viscosity, i.e. $\zeta$ vanishes~\cite{Gubser-Conformal}. In the second term of \eqref{eq:first-order} $\eta$ is the shear viscosity and $\sigma^{ab}$ is the shear tensor given by~\cite{Gourgoulhon:2006bn,Rezzolla}
	\begin{equation}\label{eq:shear-tensor}
		\sigma_{ab} = \inv{2}\left(\cd{c}{u_d}+\cd{d}{u_c}\right)\Delta^{c}_a\Delta^{d}_b-\inv{3}\Delta_{ab}\nabla_cu^c.
	\end{equation}
	In the last two terms of \eqref{eq:first-order} $q^a$ is the diffusion current given by~\cite{S3xR-Minwalla}
	\begin{equation}\label{eq:diffusion}
		q^a = -\kappa \Delta^{ab}\left(\partial_b T+a_b T\right).
	\end{equation}
	Here, $\kappa$ is the thermal conductivity, $T$ is the temperature and $a_b$ is the acceleration that reads~\cite{Rezzolla}
	\begin{equation}\label{eq:acceleration}
		a^a = u^b\nabla_b u^a.
	\end{equation}
	We assume that the fluid is symmetric under the symmetry group generated by the following Killing vectors of \eqref{eq:s3xr-metric} 
	\begin{equation}\label{eq:s3xr-symmetries}
		\pderivv{t'}\,,\qquad\pderivv{\phi'}\,,\qquad\pderivv{\chi'}.
	\end{equation}
	The first vector in \eqref{eq:s3xr-symmetries} dictates that the fluid is stationary and thus in equilibrium. To satisfy this condition any source of dissipation must vanish. Therefore, the fluid can only experience a rigid motion~\cite{S3xR-Minwalla,Gourgoulhon:2006bn}, namely
	\begin{equation*}
		u \propto \xi,    
	\end{equation*}
	in which $\xi$ is a linear combination of the Killing vectors given in \eqref{eq:s3xr-symmetries}. A choice for $\xi$ is
	\begin{equation*}
		\xi = \pderivv{t'}+\omega_1\pderivv{\phi'}+\omega_2\pderivv{\chi'}.
	\end{equation*}
	Here $\omega_1$ and $\omega_2$ are rotation parameters of the Kerr-AdS5 black hole~\cite{bir}. Therefore, using $u^2=-1$, the four-velocity is found to be
	\begin{equation}\label{eq:s3r-u}
		u^a = \frac{\gamma}{L}\left(1,0,\omega_1,\omega_2\right),\quad\text{with}\quad \gamma=(1-v^2)^{-1/2},\quad\text{and}\quad v^2 = \omega_1^2\sin^2\theta+\omega_2^2\cos^2\theta,
	\end{equation}
	It is evident that $\abs{\omega_1}<1$ and $\abs{\omega_2}<1$. Since $u$ is proportional to $\xi$ it satisfies the Killing equation~\cite{Gourgoulhon:2006bn,Rezzolla,Zee-GR-Book}
	$$
	\cd{a}{u_b}+\cd{b}{u_a}=0,
	$$ 
	and therefore the shear tensor given in \eqref{eq:shear-tensor} vanishes. The only remaining dissipative part in the energy-momentum tensor is the diffusion current given in \eqref{eq:diffusion}. By symmetries, every quantity, including acceleration, must commute with $u=u^a\partial_a$. Therefore the only non-vanishing component of the acceleration reads
	$$
	a^\theta = \christoffel{\theta}{\phi_1}{\phi_1}\left(u^{\phi_1}\right)^2+\christoffel{\theta}{\phi_2}{\phi_2}\left(u^{\phi_2}\right)^2=-\cos\theta\sin\theta\frac{\gamma^2}{L^2}\left(\omega_1^2-\omega_2^2\right).
	$$
	The following helpful relation is obtained by simple algebra
	$$
	\cos\theta\sin\theta\left(\omega_1^2-\omega_2^2\right)=\inv{2}\deriv{v^2}{\theta}=\inv{\gamma^3}\deriv{\gamma}{\theta},
	$$
	that gives rise to
	\begin{equation*}
		a^a = \inv{L^2}\left(0,-\derivv{\theta}\ln\gamma,0,0\right).
	\end{equation*}
	Plugging the above relation in \eqref{eq:diffusion} and demanding $q^a=0$ one finds
	\begin{equation}\label{eq:s3r-T}
		T = \gamma\mathcal{T},
	\end{equation}
	in which $\mathcal{T}$ is a constant. On the other hand, the conformal invariance implies
	\begin{equation*}
		p = A\mathcal{T}^4\gamma^4,\qquad \epsilon=3p. 
	\end{equation*}
	For the uncharged fluid $A$ is a constant. Plugging above and \eqref{eq:s3r-u} in \eqref{eq:fem-tensor} the energy-momentum tensor is found to be
	\begin{equation*}
		T^{ab}=\frac{A\mathcal{T}^4\gamma^6}{L^2}\begin{bmatrix}
			3 +  v^2 & 0 & 4\omega_1 & 4\omega_2 \\
			0 & 1-v^2 & 0 & 0 \\
			4\omega_1 & 0 & 3\omega_1^2+\csc^2\theta -\omega_2^2\cot^2\theta & 4\omega_1\omega_2 \\
			4\omega_2  & 0 & 4\omega_1\omega_2  & 3\omega_2^2+\sec^2\theta -\omega_2^2\tan^2\theta
		\end{bmatrix}.
	\end{equation*}
	Using
	$$
	1-v^2 = \inv{2}\left(2-\omega_1^2\left(1-\cos 2\theta\right)-\omega_2^2\left(1-\cos 2\theta\right)\right)=
	\inv{2}\left[2-\left(\omega_1^2+\omega_2^2\right)+(\omega_1^2-\omega_2^2)\cos 2\theta\right]\,,
	$$
	one finds 
	$$
	\epsilon = u_au_bT^{ab} = \frac{3\cdot 4A\mathcal{T}^4}{\left[2-\left(\omega_1^2+\omega_2^2\right)+(\omega_1^2-\omega_2^2)\cos 2\theta\right]^2}.
	$$
	We can rewrite the above relations by defining $A\mathcal{T}^4 = T_0^4/3$ as
	\begin{equation}\label{eq:s3r-epsilon}
		\epsilon = \frac{4T_0^4}{\left[2-\left(\omega_1^2+\omega_2^2\right)+(\omega_1^2-\omega_2^2)\cos 2\theta\right]^2}=T_0^4\gamma^4\,,
	\end{equation}
	and
	\begin{equation}\label{eq:s3r-fem-tensor}
		T^{ab}=\frac{T_0^4\gamma^6}{3L^2}\begin{bmatrix}
			3 +  v^2 & 0 & 4\omega_1 & 4\omega_2 \\
			0 & 1-v^2 & 0 & 0 \\
			4\omega_1 & 0 & 3\omega_1^2+\csc^2\theta -\omega_2^2\cot^2\theta & 4\omega_1\omega_2 \\
			4\omega_2  & 0 & 4\omega_1\omega_2  & 3\omega_2^2+\sec^2\theta -\omega_2^2\tan^2\theta
		\end{bmatrix}\,.
	\end{equation}
	One can check that \eqref{eq:s3r-fem-tensor} satisfies the equations of motion, namely
	$$
	\nabla_aT^{ab} = 0.
	$$
	At this point, let's compute the fluid helicity~\cite{Markakis:2016udr}. The one-form $\dform{p}=T\dform{u}$ can be assumed as the conjugate momentum of the fluid and $L=-T$ as its Lagrangian. Therefore, the fluid equations of motion, in the absence of dissipation, is obtained from
	$$
	-\dext T=\lieder{u}{\left(T\dform{u}\right)}.
	$$  
	A vorticity two-form is then given by
	\begin{equation}\label{eq:t-vort-2form}
		\dform{\Omega}=\dext\left(T\dform{u}\right).
	\end{equation}
	The above quantity is called the T-vorticity in the literature~\cite{Becattini:2015ska}. The T-vorticity vector is found from contracting $u$ with the Hodge dual of $\dform{\Omega}$, namely
	$$
	\omega^a = u_b\dual{\Omega}^{ba}.
	$$
	Plugging Eqns.~\eqref{eq:s3r-T} and \eqref{eq:s3r-u} into \eqref{eq:t-vort-2form} one finds the T-vorticity vector as 
	\begin{equation}\label{eq:s3r-tvort}
		\omega^a = \left(\frac{8}{3}\right)^{1/4}\frac{T_0\gamma^3}{L^2}\left(\omega_1\omega_2,0,\omega_2,\omega_1\right).
	\end{equation}  
	The fluid helicity is defined as~\cite{Markakis:2016udr}
	$$
	h_{f}^a = Tu_b\dual{\Omega}^{ba}.
	$$
	Putting Eqns.~\eqref{eq:s3r-T}, \eqref{eq:s3r-u}, and \eqref{eq:t-vort-2form} in above gives rise to
	\begin{equation}\label{eq:s3r-fhel}
		h^a_{f} = \left(\frac{4}{3}\right)^{1/2}\frac{T_0^2\gamma^4}{L^2}\left(\omega_1\omega_2,0,\omega_2,\omega_1\right).
	\end{equation}
	One can check that the fluid helicity is conserved, i.e. $\nabla.h=0$.
	\subsection{Passing to flat spacetime}\label{subsec:flat-hydro}
	Since the Weyl tensor of \eqref{eq:s3xr-metric} vanishes it must be locally conformally flat~\cite{Zee-GR-Book}. Motivated by the well-known conformal transformation of $S^d$ to $\mathbb{R}^d$~\cite{Zee-GR-Book}, it is found that the Weyl factor that relates $S^3\times R$ to Minkowski spacetime reads~\cite{Bantilan:2012vu}
	\begin{equation}\label{eq:conf-factor-cart}
	W=\inv{\cos t'+V/L},
	\end{equation}
	in which $V$ is defined in \eqref{eq:s3r-cartesin}. To recognize this, let's consider the ordinary parametrization of Minkowski spacetime, i.e.
	\begin{equation}\label{eq:minkowski-cart}
	\rd{s_{\mathbb{R}^{3,1}}^2}=-\rd{t^2}+\rd{x^2_1}+\rd{x^2_2}+\rd{x^2_3},
	\end{equation}	
	and use the following coordinate transformation
	\begin{equation}\label{eq:coord-trans-cart}
	t=LW\sin t',\quad x_1 = W X,\quad x_2=W Y,\quad x_3 =W Z.
	\end{equation}
	Plugging above in \eqref{eq:minkowski-cart} and using \eqref{eq:s3-eqn} gives rise to
	\begin{equation*}
	\inv{W^2}\rd{s_{\mathbb{R}^{3,1}}^2}=-L^2\rd{t'^2}+\rd{X^2}+\rd{Z^2}+\rd{Y^2}+\rd{V^2}.
	\end{equation*}
	Instead of \eqref{eq:minkowski-cart} we may parameterize the flat spacetime with the spherical coordinates defined as
	\begin{equation}\label{eq:sph-coords}
	x_1 = r\sin\theta\cos\phi,\qquad x_2 = r\sin\theta\sin\phi,\qquad x_3 = r\cos\theta. 
	\end{equation} 
	Consequently the conformal relation between $S^3\times\mathbb{R}$ and $M^4$ is found as
	\begin{equation}\label{eq:conformal-metric}
	\rd{s_{S^3\times \mathbb{R}}^2}=\inv{W^2}\left(-\rd{t^2}+\rd{r^2}+r^2\rd{\theta^2}+r^2\sin^2\theta\rd{\phi^2}\right),
	\end{equation}
	and the coordinate transformation \eqref{eq:coord-trans-cart} translates into
	\begin{eqnarray}\label{eq:coord-transformation}
		t' &=& \arctan\frac{2Lt}{L^2+r^2-t^2},\qquad
		\theta' =\arctan\frac{r\sin\theta}{\sqrt{L^2W^2-r^2\sin^2\theta}},\nonumber\\
		\phi' &=& \phi,\qquad
		\chi' = -\arctan\frac{L^2-r^2+t^2}{2Lr\cos\theta},
	\end{eqnarray} 
	while the Weyl factor reads
	\begin{equation}\label{eq:sph-weyl-factor}
	W^2\equiv \frac{L^4+(r^2-t^2)^2+2L^2(r^2+t^2)}{4L^4}.
	\end{equation}
	For further convenience, we denote the indices of the flat spacetime with greek letters $\mu, \nu, \cdots$, and the indices of $S^3\times\mathbb{R}$ with Latin letters $a, b, \cdots$. At this stage, we can transform any Weyl covariant quantity using a combination of Weyl and coordinate transformation from $S^3\times\mathbb{R}$ to the flat spacetime~\cite{Gubser-Symmetry}. Following~\cite{Gubser-Symmetry} we denote the conformal weight of a quantity with $\left[\cdots\right]$. Using~\cite{Gubser-Symmetry}
	$$
	\Big[\epsilon\Big]=4\,,
	$$
	one finds\footnote{The symbol $\gamma$ is used to denote the Lorentz factor in the $S^3\times\mathbb{R}$ spacetime.}
	\begin{equation}\label{eq:sph-epsilon-1}
	\epsilon_{\mathbb{R}^{3,1}}=\left(\frac{T_0\gamma}{W}\right)^4\,,
	\end{equation}
	or explicitly
	\begin{eqnarray}\label{eq:sph-epsilon}
	\epsilon_{\mathbb{R}^{3,1}}&=&16 L^8 T_0^4\Big[(L^4+(r^2-t^2)^2+2 L^2 t^2)(1-\omega_2^2)+2 L^2 r^2 (1-\omega_1^2)\nonumber\\&&\hspace{5cm}+2 L^2 r^2 (\omega_1^2-\omega_2^2)\cos2\theta\Big]^{-2}\,.
	\end{eqnarray}
	On the other hand, the conformal weight of $u_\mu$ is $-1$~\cite{Gubser-Symmetry}, therefore, in the language of differential forms
	$$
	{u_{\mu}}^{\!\!\!\!\mathbb{R}^{3,1}}\dext\!{x^\mu} = W {u_a}^{\!\!\!\!S^3\times\mathbb{R}} \dext\!{x^a}.
	$$
	We solve the above equation using \eqref{eq:coord-transformation}	and find the four-velocity in the flat spacetime parametrized in spherical coordinates of \eqref{eq:sph-coords}\footnote{We drop the spacetime labels unless necessary.}
	\begin{eqnarray}
	\label{eq:sph-u}
	u^t&=& \left[\frac{\epsilon_{\mathbb{R}^{3,1}}}{16L^8T_0^4}\right]^{1/4}\left(L^2+r^2+2 r t \omega_2 \cos (\theta )+t^2\right)\,,\nonumber\\
	u^r&=&\left[\frac{\epsilon_{\mathbb{R}^{3,1}}}{16L^8T_0^4}\right]^{1/4}\left(\omega_2 \cos (\theta ) \left(L^2+r^2+t^2\right)+2 r t\right)\,,\nonumber\\
	u^\theta&=&-\left[\frac{\epsilon_{\mathbb{R}^{3,1}}}{16L^8T_0^4}\right]^{1/4}\frac{\omega_2 \sin (\theta ) \left(L^2-r^2+t^2\right)}{r}\,,\nonumber\\
	u^\phi&=&2\left[\frac{\epsilon_{\mathbb{R}^{3,1}}}{16L^8T_0^4}\right]^{1/4}L\omega_1\,.
	\end{eqnarray}
	The fluid T-vorticity and helicity can also be transformed using the same method. To proceed we notice that $\Big[Tu_\mu\Big]=0$ and therefore $\Omega_{\mu\nu}$ has also zero conformal weight. Thereby,
	\begin{equation*}
	\Big[h^\mu\Big]=\Big[Tu_\nu\dual{\Omega^{\nu\mu}}\Big]=4,\qquad\Big[\omega^\mu\Big]=\Big[\dual{\Omega^{\nu\mu}u_\nu}\Big]=3.
	\end{equation*} 
	Therefore, the T-Vorticity found in \eqref{eq:s3r-tvort} is transformed into
	\begin{eqnarray*}
	\omega^t&=&\left(\inv{LT_0}\right)^2\left[\frac{\epsilon_{\mathbb{R}^{3,1}}}{27}\right]^{3/4}\frac{\omega_1 \left(\omega_2 \left(L^2+r^2+t^2\right)+2 r t \cos (\theta )\right)}{L}\,,\nonumber\\
	\omega^r&=&\left(\inv{LT_0}\right)^2\left[\frac{\epsilon_{\mathbb{R}^{3,1}}}{27}\right]^{3/4}\frac{\omega_1 \left(\cos (\theta ) \left(L^2+r^2+t^2\right)+2 r t \omega_2\right)}{L}\,,\nonumber\\
	\omega^\theta&=&-\left(\inv{LT_0}\right)^2\left[\frac{\epsilon_{\mathbb{R}^{3,1}}}{27}\right]^{3/4}\frac{\omega_1 \sin (\theta ) \left(L^2-r^2+t^2\right)}{L r}\,,\nonumber\\
	\omega^\phi&=&\left(\inv{LT_0}\right)^2\left[\frac{\epsilon_{\mathbb{R}^{3,1}}}{27}\right]^{3/4}2 \omega_2\,,
	\end{eqnarray*}
	while transformation of \eqref{eq:s3r-fhel} gives rise to
	\begin{eqnarray*}
	h_f^t&=&\frac{\epsilon_{\mathbb{R}^{3,1}}}{(LT_0)^2\sqrt{3}}\frac{\omega_1 \left(\omega_2 \left(L^2+r^2+t^2\right)+2 r t \cos (\theta )\right)}{L}\,,\nonumber\\
	h_f^r&=&\frac{\epsilon_{\mathbb{R}^{3,1}}}{(LT_0)^2\sqrt{3}}\frac{\omega_1 \left(\cos (\theta ) \left(L^2+r^2+t^2\right)+2 r t \omega_2\right)}{L}\,,\nonumber\\
	h_f^\theta&=&-\frac{\epsilon_{\mathbb{R}^{3,1}}}{(LT_0)^2\sqrt{3}}\frac{\omega_1 \sin (\theta ) \left(L^2-r^2+t^2\right)}{L r}\,,\nonumber\\
	h_f^\phi&=&\frac{\epsilon_{\mathbb{R}^{3,1}}}{(LT_0)^2\sqrt{3}}2 \omega_2\,.
	\end{eqnarray*}
	\subsection{Milne coordinates}\label{subsec:milne-hydro}
	The results of Sec.~\ref{subsec:flat-hydro} can be transformed into Milne coordinates that are related to the spherical coordinates of \eqref{eq:sph-coords} by
	\begin{eqnarray}\label{eq:milne}
	\tau&=&\sqrt{t^2-r^2\sin^2\theta\sin^2\phi}\,,\quad \eta=\inv{2}\log\frac{t+r \sin \theta\sin\phi}{t-r \sin \theta\sin\phi}\,,\nonumber\\
	x &=& r \cos \theta\,,\qquad y=r \sin\theta \cos\phi\,.
	\end{eqnarray}
	Using above equations the energy density in Milne coordinates is found to be
	\begin{eqnarray}\label{eq:milne-epsilon}
	\epsilon_{\mathbb{R}^{3,1}}&=&\frac{16\gamma^4L^8 T_0^4}{\left(L^4+2 L^2 \left(\tau ^2 \cosh (2 \eta )+{\bf x}_\perp^2\right)+\left(\tau ^2-{\bf x}_\perp^2\right)^2\right)^2}\,,
	\end{eqnarray}
	in which ${\bf x}_\perp^2=x^2+y^2$. Once again we emphasize that we use $\gamma$ as a shorthand notation for the Lorentz factor in the $S^3\times\mathbb{R}$ spacetime transformed like a scalar. The explicit form of the above formula is presented in~\cite{bir}. The components of four-velocity are given by
	\begin{eqnarray*}
	u^\tau &=& \left(\inv{LT_0}\right)\left[\frac{\epsilon_{\mathbb{R}^{3,1}}}{16L^4}\right]^{1/4}\left[(L^2+\tau^2+{\bf x}_\perp^2)\cosh \eta+2 (\tau \omega_2 x-L \omega_1 x_2 \sinh\eta)\right]\,,\nonumber\\
	u^x &=& \left(\inv{LT_0}\right)\left[\frac{\epsilon_{\mathbb{R}^{3,1}}}{16L^4}\right]^{1/4}\left[\omega_2 \left(L^2+\tau ^2+{\bf x}_\perp^2-2y^2\right)+2 \tau  x \cosh (\eta )\right]\,,\nonumber\\
	u^y &=& \left(\inv{LT_0}\right)\left[\frac{\epsilon_{\mathbb{R}^{3,1}}}{16L^4}\right]^{1/4}16\left[ -L \tau  \omega_1 \sinh (\eta )+y(x \omega_2+\tau  \cosh\eta)\right]\,,\nonumber\\
	u^\eta &=& \left(\inv{LT_0}\right)\left[\frac{\epsilon_{\mathbb{R}^{3,1}}}{16L^4}\right]^{1/4}\left[\frac{\sinh (\eta ) \left(L^2-\tau ^2+{\bf x}_\perp^2\right)-2 L y \omega_1 \cosh (\eta )}{\tau}\right]\,.
	\end{eqnarray*}
	Finally, we find the fluid helicity in Milne coordinates
	\begin{eqnarray*}
	h_f^\tau&=&\frac{\epsilon_{\mathbb{R}^{3,1}}}{(LT_0)^2\sqrt{3}}\frac{\omega_1 \omega_2 \cosh (\eta ) \left(L^2+\tau ^2+{\bf x}_\perp^2\right)-2 L y \omega_2 \sinh (\eta )+2 \tau  x \omega_1}{L}\,,\nonumber\\
	h_f^x&=&\frac{\epsilon_{\mathbb{R}^{3,1}}}{(LT_0)^2\sqrt{3}}\frac{\omega_1 \left(L^2+\tau ^2+{\bf x}_\perp^2+2 \tau  x \omega_2 \cosh (\eta )-2y^2\right)}{L}\,,\nonumber\\
	h_f^y&=&\frac{\epsilon_{\mathbb{R}^{3,1}}}{(LT_0)^2\sqrt{3}}\left[\frac{2 y \omega_1 (\tau  \omega_2 \cosh (\eta )+x)}{L}-2 \tau  \omega_2 \sinh (\eta )\right]\,,\nonumber\\
	h_f^\eta&=&-\frac{\epsilon_{\mathbb{R}^{3,1}}}{(LT_0)^2\sqrt{3}} \frac{\omega_2 \left(\omega_1 \sinh (\eta ) \left(L^2-\tau ^2+{\bf x}_\perp^2\right)-2 L y \cosh (\eta )\right)}{L \tau }\,.
	\end{eqnarray*}
	\section{Ideal MHD on the boundary}\label{sec:s3r-mhd}
	\setcounter{equation}{0}
	\par
	At this stage, we assume that the fluid on the $S^3\times\mathbb{R}$ boundary is coupled with electromagnetic fields and therefore its mutual evolution with the aforementioned fields is governed by the equation of relativistic magnetohydrodynamics (RMHD)~\cite{beckenstein1978}. The electromagnetic degrees of freedom are encoded in the field tensor, i.e.~$F_{ab}$ that satisfies the homogenous Maxwell equations
	\begin{equation}\label{eq:h-maxwell}
	\partial_a F_{bc}+\partial_b F_{ca}+\partial_c F_{ab}=0.
	\end{equation}
	Like any other quantity in the system, $F_{ab}$ must respect the symmetries expressed by \eqref{eq:s3xr-symmetries}, namely~\cite{beckenstein1978,cmhd}
	\begin{equation*}
	\partial_0F_{ab}=0\,,\qquad\partial_2F_{ab}=0\,,\qquad\partial_3F_{ab}=0\,.
	\end{equation*}
	Using above relations in \eqref{eq:h-maxwell} gives rise to
	\begin{eqnarray*}
		0&=&\partial_2F_{01}+\partial_1F_{20,1}+\partial_0F_{12}\implies\partial_1 F_{20}=0 \implies F_{20}=\text{const.}\,,\\
		0&=&\partial_3 F_{01}+\partial_1 F_{30}+\partial_0 F_{13}\implies \partial_1 F_{30}=0 \implies F_{30}=\text{const.}\,,\\
		0&=&\partial_3F_{12}+\partial_2F_{31,2}+\partial_1F_{23}\implies\partial_1 F_{23}=0 \implies F_{23}=\text{const.}
	\end{eqnarray*}
	If the gauge potential $\dform{A}$ obeys the same spacetime symmetries as other objects, then the above constants must be zero, because none of the corresponding components contain a $\theta'$-derivative. For example
	$$
	F_{20}=\partial_{\phi'}A_{t'}-\partial_{t'}A_{\phi'}=0\,.
	$$
	 Therefore
	\begin{equation*}
		F_{03}=0\,,\qquad F_{23}=0\,,\qquad F_{02}=0\,.
	\end{equation*}
	Up to now, we have only utilized the symmetry arguments. To go further, we assume that the fluid is highly conductive and employ the ideal MHD limit~\cite{beckenstein1978,cmhd}, i.e. $u^a F_{ab}=0$, that leads to
	\begin{equation*}
		F_{0a}=-\omega_1 F_{2a}-\omega_2 F_{3a}\,.
	\end{equation*}
	In particular
	\begin{equation*} 
		F_{01}=\omega_1 F_{12}-\omega_2 F_{31}\,.
	\end{equation*}
	For simplicity we define
	\begin{equation}\label{eq:scaling-funcs}
		F_{12}=2LT_0\Sigma_0f(\theta')\,,\qquad F_{31}=2LT_0\Sigma_0T_0^3h(\theta')\,,
	\end{equation} 
	in which $\Sigma_0$ is an integration constant. The field tensor is now determined up to two unknown functions 
	\begin{equation}\label{eq:s3r-farad}
		F_{ab}=2LT_0\begin{bmatrix}
			0 & \omega_1 f(\theta')-\omega_2 h(\theta')\hspace{1em} & \hspace{1em} 0 & 0 \\
			-\omega_1 f(\theta')+\omega_2 h(\theta') & 0 & f(\theta')\hspace{1em} & -h(\theta')\\
			0 & -f(\theta') & 0 & 0 \\
			0 & h(\theta') & 0 & 0
		\end{bmatrix}\,.
	\end{equation}
	\subsection{Force-free condition}
	Since we assumed the ideal MHD limit, the inhomogeneous Maxwell equations cannot help us in finding a concrete solution. Therefore the only remaining set of equations is the energy-momentum conservation that in the presence of electromagnetic fields is modified as~\cite{cmhd}
	\begin{equation}\label{eq:em-cons}
		\cd{a}{T^{ab}_{\text{fluid}}}=F^{bc}J_c\,.
	\end{equation}
	As in relativistic hydrodynamics~\cite{Gourgoulhon:2006bn}, the above equation can be decomposed into longitudinal and transverse parts with respect to the fluid velocity. In the longitudinal direction, the RHS vanishes and thus the electromagnetic degrees of freedom decouple from the energy equation~\cite{RPRR-Bjorken-iMHD,cmhd}. To obtain a solution from the transverse direction of \eqref{eq:em-cons}, we make the following assumptions 
	\begin{enumerate}
		\item \label{force-free} The fluid kinematics, namely its velocity and acceleration, is not disturbed by the electromagnetic fields.
		\item \label{neutrality} If the fluid is supposed to be highly conductive, then the local charge density given by $\rho_e=-u.J$ is relaxed quickly and the fluid is neutral.
	\end{enumerate}
	Assumption \ref{force-free}, i.e. the force-free condition,  requires that $\Delta_{cb}\cd{a}{T^{ab}_{\text{fluid}}}=0$, and therefore
	\begin{equation}
	\label{eq:force-free-original}
	\Delta_{cb}F^{bc}J_c=0\,.
	\end{equation}
	In the iMHD limit, i.e. $u_aF^{ab}=0$, we have~\cite{beckenstein1978}
	\begin{equation}\label{eq:imhd-decomp}
	F^{ab}=\epsilon^{abcd}B_c u_d\,,
	\end{equation}
	in which $B_c$ is the called the magnetic four-vector. Plugging above in \eqref{eq:force-free-original} gives rise to
	\begin{equation}\label{eq:force-free}
		f^a\equiv\epsilon^{abcd}u_b J_c B_d =0\,,
	\end{equation} 
	which is the covariant counter-part of $\boldsymbol{J}\times\boldsymbol{B}=0$. 
	The magnetic four-vector is found from $B^a = \inv{2}\epsilon^{abcd}u_b F_{cd}$ as
	\begin{eqnarray*}
		B^0&=& -\epsilon^{0123}\left(g_{\phi'\phi'}u^2 F_{13}-g_{\chi'\chi'}u^3 F_{12}\right)=-\frac{\gamma}{L^3\sin\theta'\cos\theta' }\left(\omega_1\sin^2\theta'  F_{13}-\omega_2\cos^2\theta' F_{12}\right)\,,\\
		B^1&=& -\epsilon^{0123}\left(g_{t't'}u^0 F_{23}-g_{\phi'\phi'}u^2F_{03}+g_{\chi'\chi'}u^3F_{02}\right)=0\,,\\
		B^2&=&\epsilon^{0123}\left(g_{t't'}u^0F_{13}+g_{\chi'\chi'}u^3F_{01}\right)=\frac{\gamma}{L^3\sin\theta'\cos\theta'}\left(-F_{13}+\omega_2\cos^2\theta' F_{01}\right)\,,\\
		B^3&=& -\epsilon^{0123}\left(g_{t't'}u^0F_{12}+g_{\phi'\phi'}u^2F_{01}\right)=-\frac{\gamma}{L^3\sin\theta'\cos\theta'}\left(-F_{12}+\omega_1\sin^2\theta' F_{01}\right)\,.	
	\end{eqnarray*}
	Putting \eqref{eq:s3r-farad} in above gives rise to
	\begin{eqnarray}\label{eq:s3r-bfour-components}
		B^0 &=& \frac{2T_0L\Sigma_0\gamma}{L^3\sin\theta'\cos\theta'}\left(\omega_2\cos^2\theta' f(\theta')+\omega_1\sin^2\theta' h(\theta')\right)\,,\nonumber\\
		B^1 &=& 0\,,\nonumber\\
		B^2 &=& \frac{2T_0L\Sigma_0\gamma}{L^3\sin\theta'\cos\theta'}\left[h(
		\theta')\left(1-\omega_2^2\cos^2\theta'\right)+\omega_1\omega_2\cos^2\theta' f(\theta')\right]\,,\nonumber\\
		B^3 &=& \frac{2T_0L\Sigma_0\gamma}{L^3\sin\theta'\cos\theta'}\left[f(\theta')\left(1-\omega_1^2\sin^2\theta'\right)+\omega_1\omega_2\sin^2\theta' h(\theta')\right]\,.
	\end{eqnarray}
	Generally, the inhomogeneous Maxwell equations, i.e. 
	\begin{eqnarray}\label{eq:ih-maxwell}
	\cd{b}{F^{ab}}=J^a\,,
	\end{eqnarray}
	are employed to obtain the field tensor. However, in the iMHD limit, the current is ambiguous and one cannot solve the aforementioned equations to determine the field tensor. On the other hand, for a given field tensor, inhomogeneous equations can be utilized to find the current. In the present case, we have
	\begin{equation}\label{eq:s3r-current-maxwell}
		J^a = \inv{\sqrt{-g}}\partial_b\left(F^{ab}\sqrt{-g}\right)=
		-\left(\tan\theta'-\cot\theta'\right)F^{a 1}+\pderivv{\theta'}\left(F^{a 1}\right)\,.
	\end{equation}
	Here \eqref{eq:s3xr-symmetries} is used. Raising the indices of \eqref{eq:s3r-farad} leads to
	\begin{equation*}
		F^{ab}=\frac{2T_0\Sigma_0}{L^3}\begin{bmatrix}
			0 & -\omega_1 f(\theta')+\omega_2 h(\theta')\hspace{1em} & 0 \hspace{1em} & 0 \\
			\omega_1 f(\theta')-\omega_2 h(\theta') & 0 & f(\theta')\csc^2\theta'\hspace{1em} & -h(\theta')\sec^2\theta'\\
			0 & -f(\theta')\csc^2\theta' & 0 & 0\\
			0 & h(\theta')\sec^2\theta' & 0 & 0
		\end{bmatrix}\,.
	\end{equation*}
	Plugging above in \eqref{eq:s3r-current-maxwell} we find
	\begin{eqnarray}\label{eq:s3r-jfour-components}
		J^0&=&\frac{2T_0\Sigma_0}{L^3}\left[\left(\tan\theta'-\cot\theta'\right)\left(\omega_1f(\theta')-\omega_2 h(\theta')\right)-\derivv{\theta'}\left(\omega_1f(\theta')-\omega_2 h(\theta')\right)\right]\,,\nonumber\\
		J^1&=&0\,,\nonumber\\
		J^2&=&\frac{2T_0\Sigma_0}{L^3}\left[\left(\tan\theta'-\cot\theta'\right)\frac{f(\theta')}{\sin^2\theta'}-\derivv{\theta'}\left(\frac{f(\theta')}{\sin^2\theta'}\right)\right]\,,\nonumber\\
		J^3&=&-\frac{2T_0\Sigma_0}{L^3}\left[\left(\tan\theta'-\cot\theta'\right)\frac{h(\theta')}{\cos^2\theta'}-\derivv{\theta'}\left(\frac{h(\theta')}{\cos^2\theta'}\right)\right]\,.
	\end{eqnarray}
	With a bit of algebra, the following relation is obtained from \eqref{eq:s3r-jfour-components}
	\begin{equation}\label{eq:currents-aux}
		J^0 = \omega_1 \sin^2\theta' J^2 + \omega_2 \cos^2\theta' J^3 - 2\frac{2T_0\Sigma_0}{L^3}\left[\omega_1\cot\theta' f(\theta')+\omega_2\tan\theta' h(\theta')\right]\,.
	\end{equation}
	Using $J^1=0$ the components of $f^a$ in \eqref{eq:force-free} read
	\begin{eqnarray*}
		f_0&=& \sqrt{-g}B^1\gamma L\left(\omega_1 J^3-\omega_2 J^2\right)\,,\nonumber\\
		f_1&=&\sqrt{-g}\gamma L\left[B^3\left(J^2-\omega_1 J^0\right)-B^2\left(J^3-\omega_2 J^0\right)+B^0\left(\omega_1J^3-\omega_2J^2\right)\right]\,,\nonumber\\
		f_2 &=& \sqrt{-g}\gamma L B^1 \left(J^3-\omega_2 J^0\right)\sin^2\theta'\,,\nonumber\\
		f_3&=&\sqrt{-g}\gamma L B^1 \left(\omega_1J^0-J^2\right)\cos^2\theta'\,.
	\end{eqnarray*}
	Thus, the force-free condition \eqref{eq:force-free} requires that
	\begin{equation}\label{eq:force-free-nzalpha}
		\omega_2 J^2 = \omega_1 J^3\,,\quad J^3=\omega_2 J^0\,, \quad J^2=\omega_1J^0\,.
	\end{equation}
	\subsection{Current-free solution}
	At this stage, we employ the assumption of fluid's neutrality. Using $J^1=0$ and \eqref{eq:s3r-u} in $\rho_e=-u.J$, gives rise to
	$$
	\rho_e = \frac{\gamma}{L^3}J^0\left(1-\omega_1^2\csc^2\theta'-\omega_2^2\sec^2\theta'\right)\,.
	$$
	Therefore $\rho_e=0$ requires that $J^0=0$. Putting the latter result in \eqref{eq:force-free-nzalpha} shows that a vanishing local charge density leads to a free-current condition, i.e. $J^a=0$. According to \eqref{eq:s3r-jfour-components}, the two azimuthal components of the four-current vanish if 
	$$
	f(\theta')=C_1\tan\theta',\qquad h(\theta')=C_2\cot\theta'\,,
	$$
	in which $C_1$ and $C_2$ are some constants. Plugging above in $J^0$ from \eqref{eq:s3r-jfour-components} or \eqref{eq:currents-aux} gives rise to $C_2=-\omega_1C_1/\omega_2$. By some rearrangement, functions $f$ and $h$ can be written as
	\begin{equation}\label{eq:cfsol}
		f(\theta')=\beta\omega_2\frac{\tan\theta'}{2}\,,\qquad h(\theta')=-\beta\omega_1\frac{\cot\theta'}{2}\,,
	\end{equation}
	in which $\beta$ is a constant. At this stage the equations of RMHD are completely solved on $S^3\times\mathbb{R}$. Plugging \eqref{eq:cfsol} in \eqref{eq:s3r-bfour-components}, we find the magnetic four-vector 
	\begin{equation}\label{eq:s3r-fourb}
		\hspace{-1cm}B^a = \frac{T_0\gamma}{L^2}\Sigma_0\Big(-(\omega_1^2-\omega_2^2),0,-(1-\omega_2^2)\omega_1\csc^2\theta',-(1-\omega_1^2)\omega_2\sec^2\theta'\Big).
	\end{equation}
	Here $\beta$ is absorbed into $\Sigma_0$. The magnitude of $B^a$ reads
	\begin{equation}
		B=\frac{2T_0\gamma}{L\sin 2\theta'}\Sigma_0\sqrt{\omega_1^2\sin^2\theta+\omega_2^2\cos^2\theta-\omega_1^2\omega_2^2}.
	\end{equation}	
	As the above equation suggests, no magnetic field is produced if there is no rotation. Also, the magnetic field blows up both at the north pole, i.e.~$\theta'=\pi/2$ and the equator, i.e.~$\theta'=0$, if both angular velocities are non-zero. However, each of the two angular parameters is related to one of the aforementioned singularities. If $\omega_1$($\omega_2$) is absent then the magnetic field becomes regular at the pole(equator)
	$$
	\lim_{\theta'\to 0} B\at{\omega_1=0}=\frac{T_0\Sigma_0}{L}\omega_2\,,\qquad\lim_{\theta'\to \pi/2} B\at{\omega_2=0}=\frac{T_0\Sigma_0}{L}\omega_1\,.
	$$
	The plasma sigma~\cite{cmhd}, i.e. the ratio of the magnetic energy density to fluid's one, reads
	\begin{equation}\label{eq:s3r-sigma}
	\sigma = \left(\frac{\Sigma_0(1-v^2)}{LT_0\sin\theta'\cos\theta'}\right)^2\zeta\,,
	\end{equation}
	in which 	
	\begin{equation*}
	2\zeta\equiv (1-v^2)-(1-\omega_1^2)(1-\omega_2^2)\,.
	\end{equation*}
	Using \eqref{eq:s3r-sigma}, we find the Alfven speed to be~\cite{gedalin1993}
	\begin{equation}\label{eq:s3r-alfven}
	v_A=\frac{\Sigma_0}{L T_0}\sqrt{\frac{6 \zeta }{ \left(\gamma^2 \sin (2\theta')\right)^2+6 \zeta  \left(\Sigma_0/(L T_0)\right)^2}}\,.
	\end{equation}
	Both at the equator and at the north pole, the Aflven speed $v_A$ tends to unity, which is twice smaller than required by causality~\cite{gedalin1993}. Although the causality is guaranteed, one may argue that such a large Aflven speed is a sign that the force-free condition is a poor approximation close to the pole and the equator. For angles other than $0$ and $\pi/2$ the force-free condition is consistent, i.e. $v_a \ll c_s$, if 
	$$
	\Sigma_0 \ll L T_0\,.
	$$	
	One should bear in mind that the above relation is sufficient but not necessary. For $\Sigma_0 \ll L T_0$, the Alfven speed can be written as
	\begin{eqnarray}\label{eq:s3r-alfven-series}
	v_A=\frac{\Sigma_0}{L T_0}\frac{\sqrt{6\zeta}}{\gamma^2 \sin (2\theta')}+\orderof{\left(\frac{\Sigma_0^3}{L^3 T_0^3}\right)}\,.
	\end{eqnarray} 
	The electromagnetic part of the energy-momentum tensor is~\cite{beckenstein1978}
	\begin{equation*}
	T^{ab}_{\mbox{\tiny{EM}}} = F^a_{~c} F^{bc} - \inv{4}g^{ab}F_{cd}F^{cd}\,.
	\end{equation*}
	Plugging the solution \eqref{eq:cfsol} in \eqref{eq:s3r-farad} and using the result in above gives rise to
	\begin{equation*}
	T^{ab}_{em}=\frac{4T_0^2\Sigma_0^2}{L^4\sin^2 2\theta'}
	\left(
	\begin{array}{cccc}
	\zeta+\omega_1^2\omega_2^2 & 0 & \omega_1 \omega_2^2 & \omega_1^2 \omega_2 \\
	0 & \zeta & 0 & 0 \\
	\omega_1 \omega_2^2  & 0 &  \omega_2^2-\zeta\csc ^2(\theta') & \omega_1 \omega_2  \\
	\omega_1^2 \omega_2 & 0 & \omega_1 \omega_2  & \omega_1^2-\zeta\sec ^2(\theta') \\
	\end{array}
	\right)\,.
	\end{equation*}
	One can check that the divergence of the above energy-momentum tensor vanishes and therefore our solution satisfies all equations of motion. 
	\subsection{Gauge potential and magnetic helicity}
    We close this section by computing the magnetic helicity which its covariant definition is~\cite{Markakis:2016udr} 
    \begin{equation*}
    h_{em}^a = \dual{F^{ab}}A_b\,.
    \end{equation*}
    Here $A_b$ is the gauge potential and $\dual{F^{ab}}$ is the dual of the field tensor. To find $A_b$ we need to solve the equation $\dform{F}=\dext\!\dform{A}$. As a subsequent of the symmetries represented in \eqref{eq:s3xr-symmetries}, the $\theta'$ component of the gauge potential is arbitrary and using the gauge freedom we assume it to be zero. Thus the gauge one-form reads 
    \begin{equation*}
    \dform{A}=A_0(\theta')\dext\!{t'}+A_2(\theta')\dext\!{\phi'}+A_3(\theta')\dext\!{\chi'}\,.
    \end{equation*}
     Plugging above in $\dform{F}=\dext\!\dform{A}$ and using \eqref{eq:s3r-farad} gives rise to
     \begin{equation*}
     \dform{A}= LT_0\Sigma_0\left(\omega_1\omega_2 \log\cot\theta'\dext\!{t'}-\omega_2 \log\cos\theta'\dext\!{\phi'}+\omega_1 \log\sin\theta'\dext\!{\chi'}\right)\,.
     \end{equation*}  
      To find $\dfdual{F}$ we use $\dfdual{F}=-\dform{u}\wedge\dform{B}$, which is the Hodge star dual of \eqref{eq:imhd-decomp} in the language of differential forms. The result is 
      \begin{equation}
      \dfdual{F}=-LT_0\Sigma_0\left(\omega_1\dext\!t'\wedge\dext\!\phi'-\omega_2\dext\!t'\wedge\dext\!\chi'+\omega_1\omega_2\dext\!\phi'\wedge\dext\!\chi'\right)\,.
      \end{equation}
      Contracting $\dform{A}$ with above we find the magnetic helicity to be $$
      {h^a}_{em}=\frac{T0^2\omega_1^2\omega_2^2}{L^2}\Sigma_0^2\left(\frac{\log\cos\theta'}{\sin^2\theta'}+\frac{\log\sin\theta'}{\cos^2\theta'}\right)\left(1,0,\omega_1,\omega_2\right)\,,
      $$ 
      or
      \begin{equation}\label{eq:s3r-hm}
      \dform{h}=\frac{T0^3\omega_1^2\omega_2^2}{L}\Sigma_0^2\left(\frac{\log\cos\theta'}{\sin^2\theta'}+\frac{\log\sin\theta'}{\cos^2\theta'}\right)\,\epsilon^{-1/4}\dform{u}\,.
      \end{equation}
       As anticipated the magnetic helicity is conserved, i.e. $\nabla.h_{em}=0$. Also, a non-vanishing magnetic helicity requires both angular parameters to be non-zero.
	\section{Passing to flat spacetime}
	\label{sec:flat-imhd}
	\setcounter{equation}{0}
	\par
	Now we are in a situation to find the solution to the equations of RMHD in the flat spacetime. The procedure is similar to Sec.~\ref{sec:bir-hydro} and is previously examined in~\cite{cmhd}.  As in Sec.~\ref{sec:bir-hydro} and~\cite{bir}, the first step is to perform a combination of a Weyl and a coordinate transformation to the spherical coordinates given in \eqref{eq:conformal-metric}. The magnetic differential form has a conformal weight of $1$~\cite{cmhd} thus 
	$$
	{B_{\mu}}^{\!\!\!\!\mathbb{R}^{3,1}}\dext\!{x^\mu} = \inv{W}{B_a}^{\!\!\!\!S^3\times\mathbb{R}} \dext\!{x^a}\,.
	$$
	 Using \eqref{eq:s3r-fourb} and \eqref{eq:sph-epsilon-1} the above relation is transformed into 
	 \begin{equation*}
	 {B_{\mu}}^{\!\!\!\!\mathbb{R}^{3,1}} \dext\!{x^\mu} =\epsilon_{\mathbb{R}^{3,1}}^{1/4}\,\Sigma_0\left((\omega_1^2-\omega_2^2)\dext\!t'-\omega_1(1-\omega_2^2)\dext\!\phi'+\omega_2(1-\omega_1^2)\dext\!\chi'\right)\,.
	 \end{equation*} 
	 We then use the coordinate transformations of \eqref{eq:coord-transformation} in above and solve the resulting equation to obtain $B^\mu$. This process gives rise to
	 \begin{eqnarray*}
	 B^t &=& -\Sigma_0\,\left[\frac{\epsilon}{L^4}\right]^{1/4}\left[\frac{\left(\omega _1^2-\omega_2^2\right) \left(L^2+r^2+t^2\right)}{2 L^2 W^2 }-\frac{rt\left(1-\omega _1^2\right)\omega_2\cos\theta}{L^2 W^2-r^2 \sin ^2(\theta )}\right]\,,\nonumber\\
	 B^r &=& \Sigma_0\,\left[\frac{\epsilon}{L^4}\right]^{1/4}\left[\frac{\left(1-\omega_1^2\right) \omega_2 \cos\theta\left(L^2+r^2+t^2\right)}{2\left(L^2 W^2-r^2\sin ^2\theta \right)}-\frac{2rt\left(\omega _1^2-\omega_2^2\right)}{2 L^2 W^2}\right]\,,\nonumber\\
	 B^\theta &=& -\Sigma_0\,\left[\frac{\epsilon}{L^4}\right]^{1/4}\left[\frac{\left(1-\omega_1^2\right)\omega_2 \left(L^2-r^2+t^2\right)}{L^2 W^2-r^2\sin ^2\theta }\right]\frac{\sin (\theta )}{2r}\,, \nonumber\\
	 B^\phi &=& -\Sigma_0\,\left[\frac{\epsilon}{L^4}\right]^{1/4} \omega_1\left(1-\omega_2^2\right)\frac{L}{r\sin\theta}. 
	 \end{eqnarray*}
	  The magnitude of $B$ is also obtained to be 
	  \begin{eqnarray*}
	  B &=& \Sigma_0\,\left(\frac{\epsilon^{1/4}}{LW}\right)\Bigg[L^2W^2\left(\frac{\omega_1^2(1-\omega_2^2)}{r^2\sin^2\theta}+\frac{\omega_2^2(1-\omega_1^2)}{L^2W^2-r^2\sin^2\theta}\right)-(\omega_1^2-\omega_2^2)\Bigg].
	  \end{eqnarray*}
	   At this stage, we transform the results to the Milne coordinates using \eqref{eq:milne}. Applying the ordinary coordinate transformation~\cite{Zee-GR-Book}, the magnetic four-vector in the Milne coordinates is found to be 
	   \begin{eqnarray*}
	   B^\tau &=& \frac{\Sigma_0}{L}\Bigg[\frac{y \omega_1 \left(1-\omega_2^2\right)L\sinh \eta}{\tau ^2 \sinh ^2\eta +y^2}+\frac{x\tau \left(1-\omega_1^2\right)\omega_2}{L^2 W^2-\tau ^2 \sinh ^2\eta-y^2 }-\frac{\cosh (\eta ) \left(\omega_1^2-\omega_2^2\right) \left(L^2+\tau ^2+{\bf x}_\perp^2\right)}{2L^2 W^2}\Bigg]\epsilon^{1/4}\,,\nonumber\\
	   B^x &=&\frac{\Sigma_0}{L}\Bigg[\frac{\left(1-\omega_1^2\right) \omega_2 \left(L^2+\tau ^2+x^2-y^2\right)}{2\left(L^2 W^2-\tau ^2 \sinh ^2\eta -y^2\right)}-\frac{ x\tau\cosh \eta \left(\omega_1^2-\omega_2^2\right)}{L^2 W^2}\Bigg]\epsilon^{1/4}\,,\nonumber\\
	   B^y &=& \frac{\Sigma_0}{L}\Bigg[\frac{x y \left(1-\omega_1^2\right) \omega_2}{L^2 W^2-\tau ^2 \sinh ^2\eta-y^2}-\frac{ y\tau\cosh\eta \left(\omega_1^2-\omega_2^2\right)}{L^2 W^2}+\frac{L \tau  \omega_1 \left(1-\omega_2^2\right) \sinh\eta}{\tau ^2\sinh ^2\eta+y^2}\Bigg]\epsilon^{1/4}\,,\nonumber\\
	   B^\eta &=& \frac{\Sigma_0}{L\tau}\Bigg[\frac{\sinh (\eta ) \left(\omega_1^2-\omega_2^2\right) \left(L^2-\tau ^2+{\bf x}_\perp^2\right)}{2 L^2 W^2}-\frac{L y \omega_1 \left(1-\omega_2^2\right) \cosh (\eta )}{\tau ^2 \sinh ^2(\eta )+y^2}\Bigg]\epsilon^{1/4}\,.
	   \end{eqnarray*}
	   Here $W$ and $\epsilon$ are supposed to be expressed in Milne coordinates. One can verify that the above results satisfy all equations of motion. The behavior of the solution at $\tvec{x}=0$, i.e. $x=0\,,y=0\,,$ and $\eta=0$ is as follows: The $\tau$ and $x$ components are nonzero and finite even at $\tau=0$. The first component vanishes if $\omega_1=\omega_2$ and the second if $\omega_2=0$. The $y$ component is zero at the aforementioned point no matter what is the value of $\tau$.	The $\eta$ component is not regular at $y=0$ and behaves as
		$$
		\tau B^\eta \propto \inv{y/L}\,.
		$$
	This singular behavior is translated to $B$ and is related to the singularity discussed in Sec.~\ref{sec:s3r-mhd}. The magnitude of the magnetic field in Milne coordinates is found to be
	\begin{eqnarray}\label{eq:milne-b}
		B&=&\frac{\Sigma_0}{L}\Bigg[\left(\frac{x\tau\cosh\eta \left(\omega_1^2-\omega_2^2\right)}{L^2 W^2}-\frac{\left(1-\omega_1^2\right) \omega_2}{2}\frac{L^2+\tau^2+x^2-y^2}{L^2 W^2-\tau ^2 \sinh^2\eta-y^2}\right)^2\\&&+\left(\frac{\omega_1^2-\omega_2^2}{2}\frac{\left(L^2-\tau^2+{\bf x}_\perp^2\right)\sinh\eta}{L^2W^2}-\frac{\omega_1(1-\omega_2^2)Ly\cosh\eta}{y^2+\tau^2\sinh^2\eta}\right)^2\nonumber\\&&+\left(\frac{\omega_1^2-\omega_2^2}{2}\frac{\left(L^2+\tau^2+{\bf x}_\perp^2\right)\cosh\eta}{L^2W^2}-\frac{\omega_2(1-\omega_1^2)x\tau}{L^2W^2-\tau^2\sinh^2\eta-y^2}-\frac{\omega_1(1-\omega_2^2)Ly\sinh^2\eta}{\tau^2\sinh^2\eta+y^2}\right)^2\nonumber\\&&+\left(\frac{y\tau(\omega_1^2-\omega_2^2)\cosh\eta}{L^2W^2}-\frac{\omega_1(1-\omega_2^2)L\tau\sinh\eta}{\tau^2\sinh^2\eta+y^2}-\frac{\omega_2(1-\omega_1^2)xy}{L^2W^2-\tau^2\sinh^2\eta-y^2}\right)^2\Bigg]^{1/2}\epsilon^{1/4}\,.\nonumber
	\end{eqnarray}
	As in Sec.~\ref{sec:s3r-mhd} there are two singular points related to the two angular parameters. If $\omega_1=0$, then $B$ tends to infinity at $y=L$, but is regular at $y=0$. On the other hand for $\omega_1\neq 0$, the point $y=0$ is singular, but the magnetic field at $y=L$ becomes finite if one assumes $\omega_2=0$. The leading behavior of $B$ around $y/L=0$ reads	
	\begin{equation}\label{eq:singular-y0}
	B\at[\Big]{x=0\,,y\ll L\,,\eta=0}=2LT_0\Sigma_0\frac{\sqrt{\omega_1^2(1-\omega_2^2)}}{(L^2+\tau^2)}\inv{y/L}+\orderof{(y/L)}\,.
	\end{equation}
	We come back to the singularity problem in Sec.~\ref{sec:regulaization}.
	\par
	 Before closing this section, let's compute the magnetic helicity in the flat spacetime. The field differential form and consequently the gauge potential have a conformal weight of zero~\cite{cmhd}. Accordingly, the conformal weight of the magnetic helicity vector is found from 
	$$
	\Big[h_{em}^\alpha\Big]=\Big[\dual{F}^{\alpha\beta}\Big]+\Big[A_\beta\Big]=-\Big[\sqrt{-g}\Big]+\Big[F_{\alpha\beta}\Big]=4\,.
	$$
	 Consequently, the conformal weight of the magnetic helicity one-form is $2$. As for the magnetic four-vector, we apply a combined Weyl and coordinate transformation to obtain the magnetic helicity in flat spacetime parameterized with spherical coordinates of \eqref{eq:conformal-metric}. The result is
	 
	 \begin{eqnarray*}
	 h_{em}^\mu&=&\frac{L \Sigma_0^2 T_0^3 \omega_1 \omega_2}{2 r^2 W^2 \sqrt[4]{\epsilon } \sin ^2(\theta )}\Bigg[\frac{L^2 W^2}{L^2 W^2-r^2 \sin ^2\theta } \log \left(\frac{L^2 W^2-r^2 \sin ^2(\theta )}{L^2 W^2}\right)\nonumber\\&&\hspace{2cm}-\frac{r^2 \sin^2\theta}{L^2 W^2-r^2 \sin ^2\theta } \log \left(\frac{L^2 W^2-r^2 \sin ^2(\theta )}{r^2 \sin ^2(\theta )}\right)\Bigg]u^\mu\,.
	 \end{eqnarray*} 
	 By transforming above we find the magnetic helicity in Milne coordinates to be 
	 \begin{eqnarray*}
	 h_{em}^\mu&=&\frac{L \Sigma_0^2 T_0^3 \omega_1 \omega_2}{2 W^2 \sqrt[4]{\epsilon } \left[L^2W^2-(\tau^2\sinh^2\eta+y^2)\right]}\Bigg[\frac{L^2 W^2}{\tau^2\sinh^2\eta+y^2} \log \left(\frac{L^2W^2-(\tau^2\sinh^2\eta+y^2)}{L^2 W^2}\right)\nonumber\\&&\hspace{4cm}- \log \left(\frac{L^2W^2-(\tau^2\sinh^2\eta+y^2)}{\tau^2\sinh^2\eta+y^2 }\right)\Bigg]u^\mu\,.
	 \end{eqnarray*}
	\section{Regulaziation with sources}\label{sec:regulaization}
	\setcounter{equation}{0}
	\par	
	In this section, we comment on the nature of singularities and provide a physical understanding of their emergence. In Sec.~\ref{sec:s3r-mhd}, we made two assumptions in addition to the iMHD limit approximation, i.e. the force-free condition and the assumption of neutrality. These two assumptions led to a current-free solution presented in \eqref{eq:cfsol}. Although it is physically possible for an infinitely conductive fluid to freeze an initial magnetic field within itself, a vanishing electrical current is a source for singularities. This can be revealed by a closer look at \eqref{eq:singular-y0} in which the singular leading term is similar to the magnetic field produced by a point-like wire in the z-direction. Therefore the singularity may be regularized if we put point-like sources of current at the equator and pole of the $S^3\times\mathbb{R}$ spacetime. Equivalently we can start by changing functions $f$ and $h$ in Eq.~\eqref{eq:scaling-funcs}	
	\begin{equation*}
	f(\theta')=f_0(\theta')+\delta f(\theta')\,,h(\theta')=h_0(\theta')+\delta h(\theta')\,.
	\end{equation*}
	Here $f_0$ and $h_0$ are the functions presented in  the current free solution of \eqref{eq:cfsol} with $\beta=1$. Maxwell equations do not put any constraint on $f$ and $h$. Subsequently adding any differentiable function to $f$ and $h$ provides another solution to Maxwell equations. However, the force-free condition is not met unless $\delta f=\delta h=0$, and one requires to solve the energy-momentum conservation to obtain the alterations in the kinematics and thermodynamics of the fluid. For simplicity, let's consider that the disturbance of the fluid by such additions is negligible and $\delta f$ and $\delta h$ are such that the magnetic field becomes regular at the equator and north pole of $S^3\times\mathbb{R}$. To realize this purpose they should be in the following form 
	\begin{eqnarray}\label{eq:regulrization}
	\delta f(\theta')&=&f_\lambda(\theta')\left(Q_1\sin\theta'-f_0(\theta')\right)-h_\lambda(\theta')f_0(\theta')\,,\nonumber\\
	\delta h(\theta')&=&h_\lambda(\theta')\left(-Q_2\cos\theta'-h_0(\theta')\right)-f_\lambda(\theta')h_0(\theta')\,.
	\end{eqnarray}	
	Functions $f_\lambda$ and $h_\lambda$ in the above Ansatz are used to \textit{turn off} $f$ and $h$, as well as the regulators, i.e. $Q_1\sin\theta'$ and  $-Q_2\cos\theta'$ in the equator and pole. $Q_1$ and $Q_2$ are numbers to be determined by the initial and boundary conditions. In order for $B$ to be regular in $\theta'=0$ and $\theta'=\pi/2$ it is required that
	\begin{eqnarray}\label{eq:reg-conditions}
	f_\lambda(0)=1\,,\qquad\lim_{\lambda\to 0}f(\theta'>0)=0\,,\qquad h_\lambda(\pi/2)=1\,,\qquad\lim_{\lambda\to 0}f(\theta'<\pi/2)=0\,.
	\end{eqnarray}
	Here $\lambda$ is a small dimensionless parameter needed for guaranteeing the above conditions. Assuming the above conditions in disturbing the current free solution we find that the magnetic field has following finite limits at the pole and the equator
	\begin{equation}\label{eq:b-limits}
	\lim_{\theta'\to 0} B = \frac{2Q_1T_0\Sigma_0}{L}\,,\qquad
	\lim_{\theta'\to \pi/2} B = \frac{2Q_2T_0\Sigma_0}{L}.
	\end{equation}
	However, as anticipated the regulation does not come without a price. The current is not vanishing anymore and in particular, the following local charge density appears in the fluid
	\begin{equation}\label{eq:s3r-charge}
	\delta{\rho_e}^{S^3\times\mathbb{R}} = -\frac{4T_0\Sigma_0}{L^2}\left[Q_1\omega_1\cos\theta'f_\lambda(\theta')-Q_2\omega_2\sin\theta'h_\lambda(\theta')\right]\gamma.
	\end{equation}
	Also the fluid's acceleration is modified at both $\theta'=0$ and $\theta'=\pi/2$. The induced charge by the regularization is small if we carefully choose the parameters. However, the produced electromagnetic acceleration at $\theta'=0$ and $\theta'=\pi/2$ might be still large. This is because the acceleration at these points is proportional to $f'_\lambda(0)$ and $h'_\lambda(\pi/2)$, which are large by the conditions of \eqref{eq:reg-conditions}. Although the regularization proposed in \eqref{eq:regulrization} is ad hoc, it demonstrates the role of the electrical current in a consistent solution to the equations of RMHD. 
	\section{Numerical results}\label{sec:results}
	\setcounter{equation}{0}
	\par
\begin{figure*}[hbt]\centering	
		\includegraphics[width=7cm]{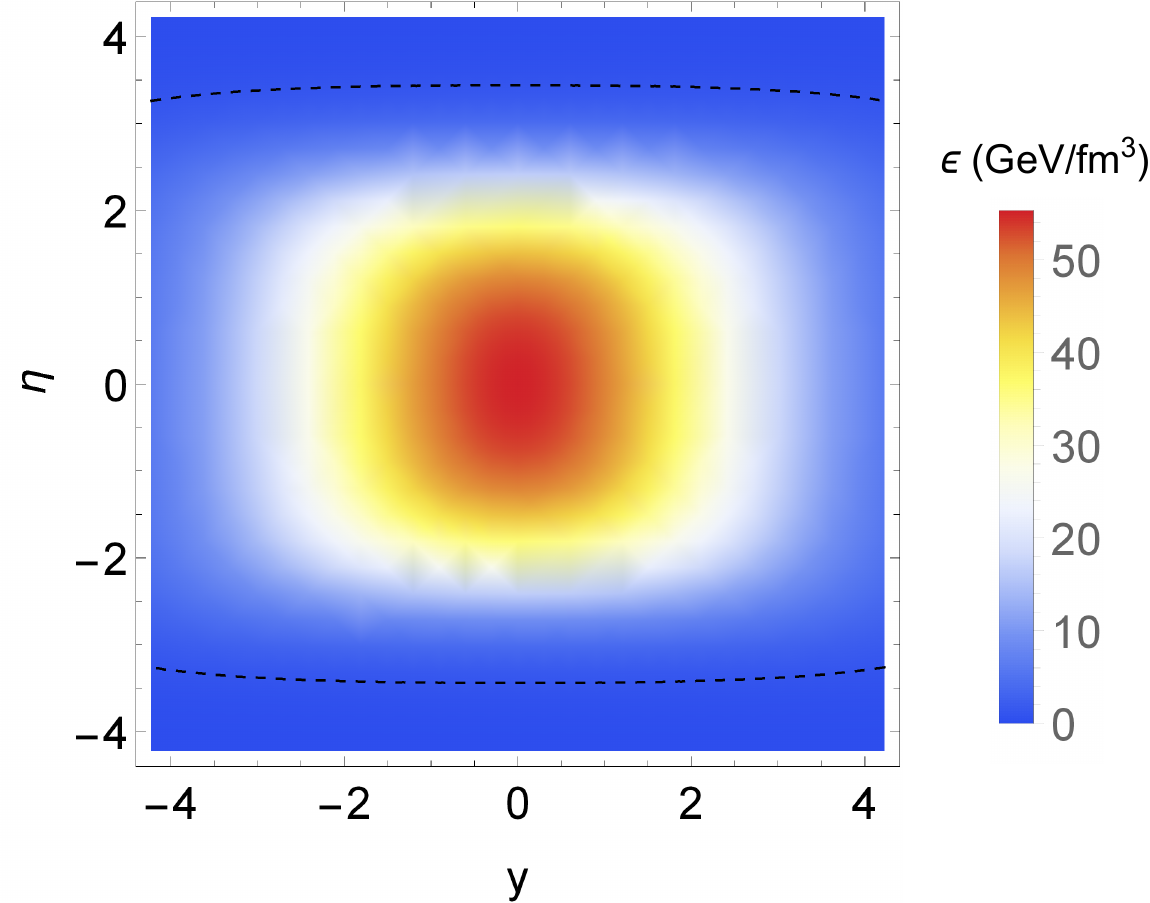}
		\hspace{0.3cm}
		\includegraphics[width=7cm]{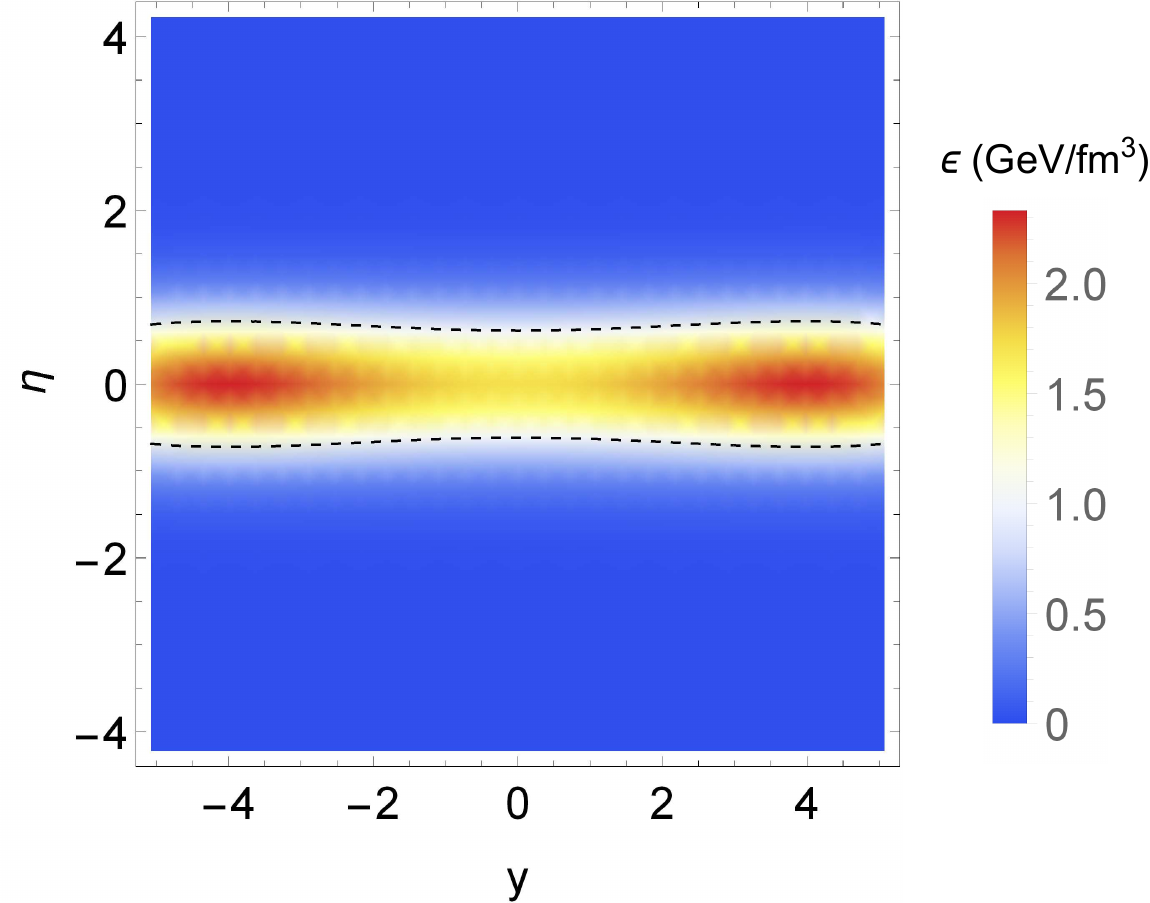}
		\caption{(color online). Energy density in $y\!-\!\eta$ plane from Eq.~\eqref{eq:milne-epsilon} using parameters fixed in \eqref{eq:params} for $\omega_1=0.5\,,\omega_2=0.05$ at (a) $\tau=0.4\fmc$ and  (b) $\tau=5\fmc$. The dashed line is the contour of $\epsilon=1~\GeV/\fm^3$. }\label{fig:epsilonyeta}
	\end{figure*}
\begin{figure*}[hbt]\centering	
		\includegraphics[width=7cm]{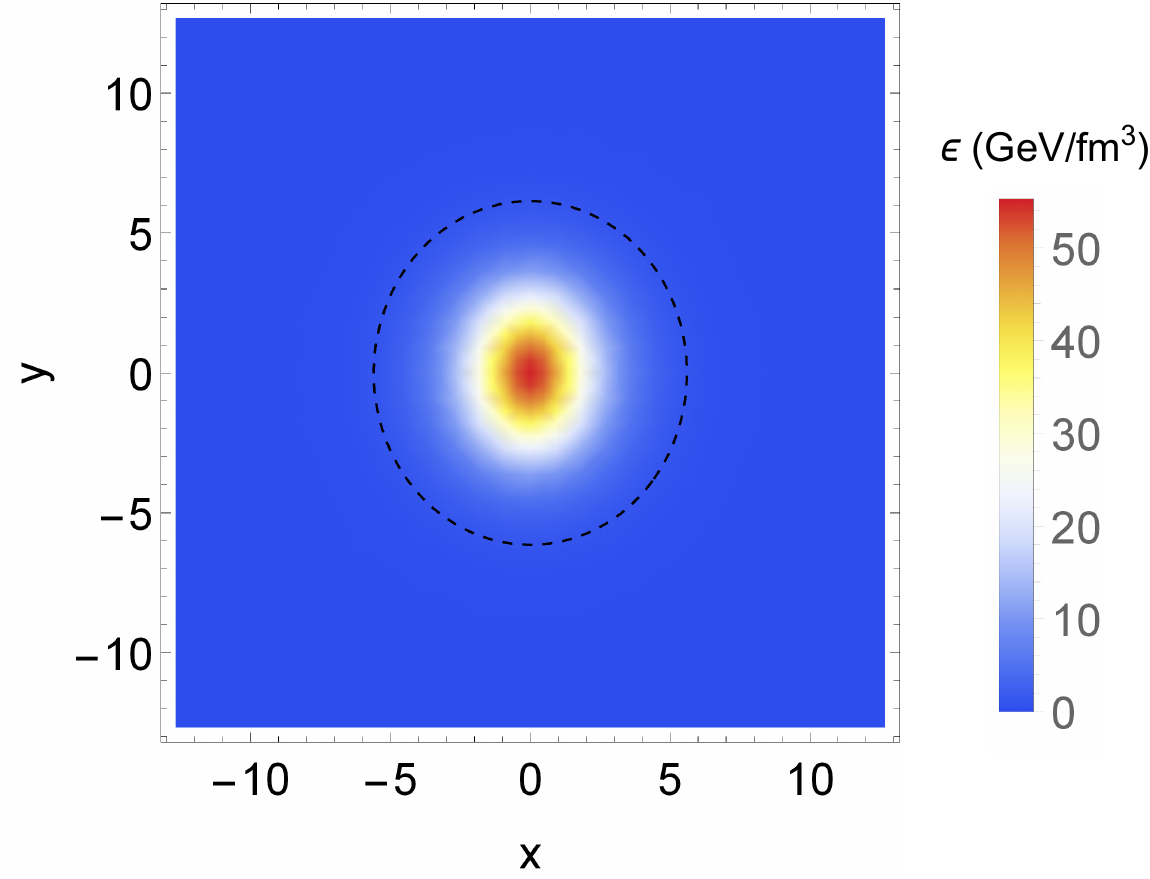}
		\hspace{0.3cm}
		\includegraphics[width=7cm]{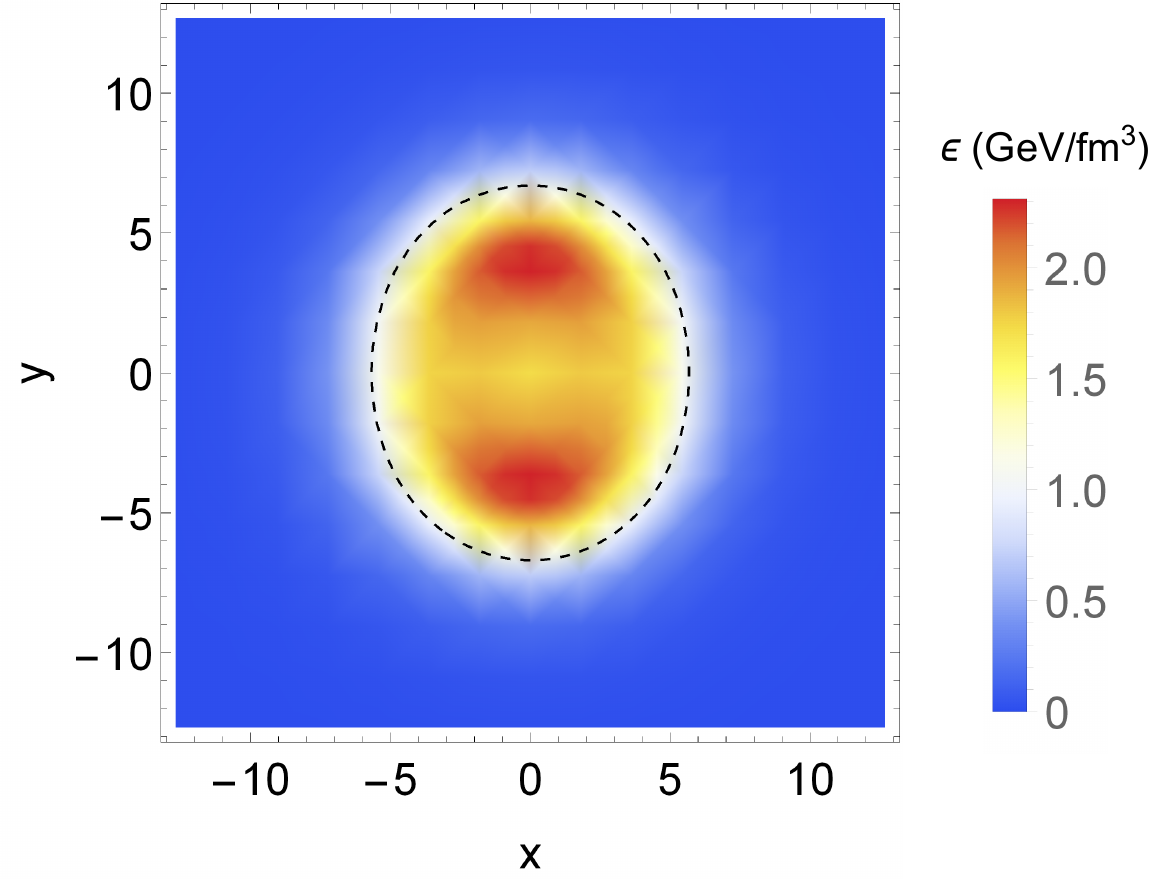}
		\caption{(color online). Energy density in the transverse plane from Eq.~\eqref{eq:milne-epsilon} using parameters fixed in \eqref{eq:params} for $\omega_1=0.5\,,\omega_2=0.05$ at (a) $\tau=0.4\fmc$ and  (b) $\tau=5\fmc$. The dashed line is the contour of $\epsilon=1~\GeV/\fm^3$. }\label{fig:epsilonxy}
	\end{figure*}
\begin{figure*}[hbt]\centering	
		\includegraphics[width=7cm]{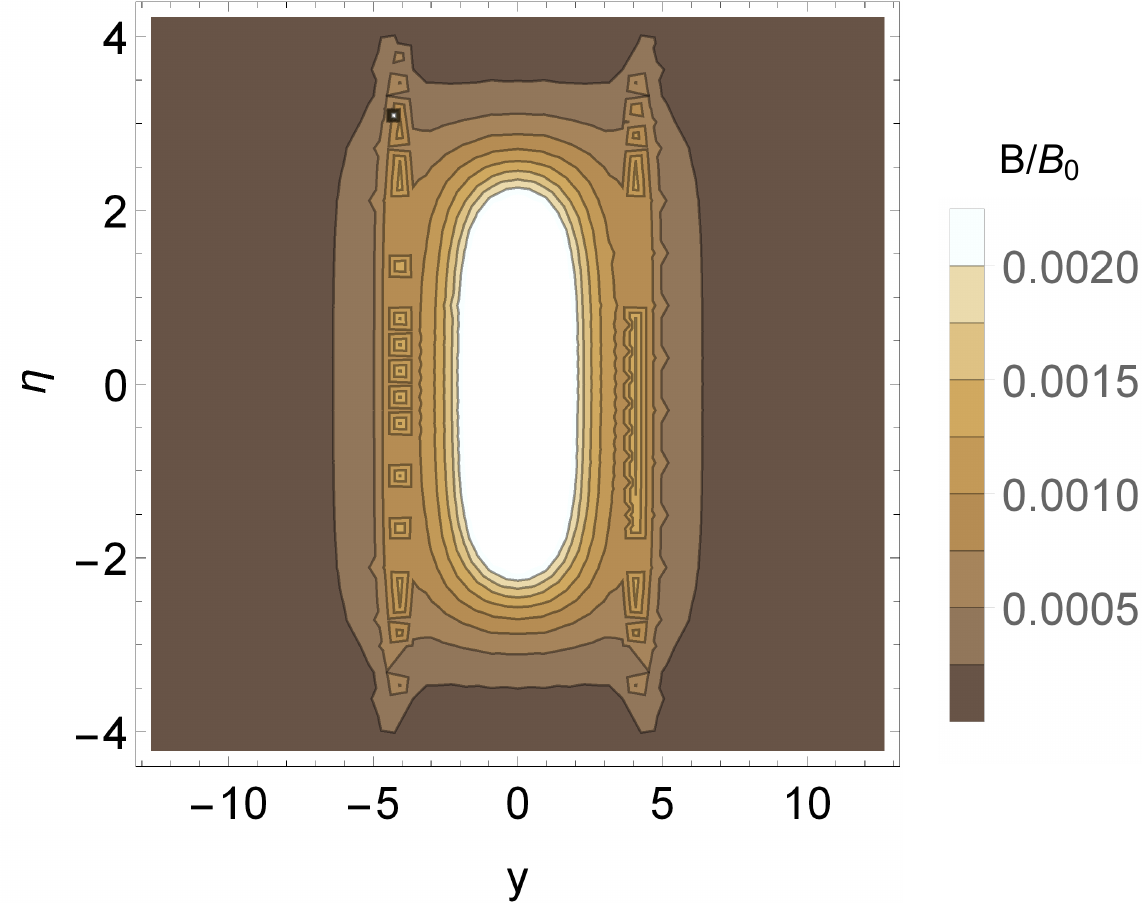}
		\hspace{0.3cm}
		\includegraphics[width=7cm]{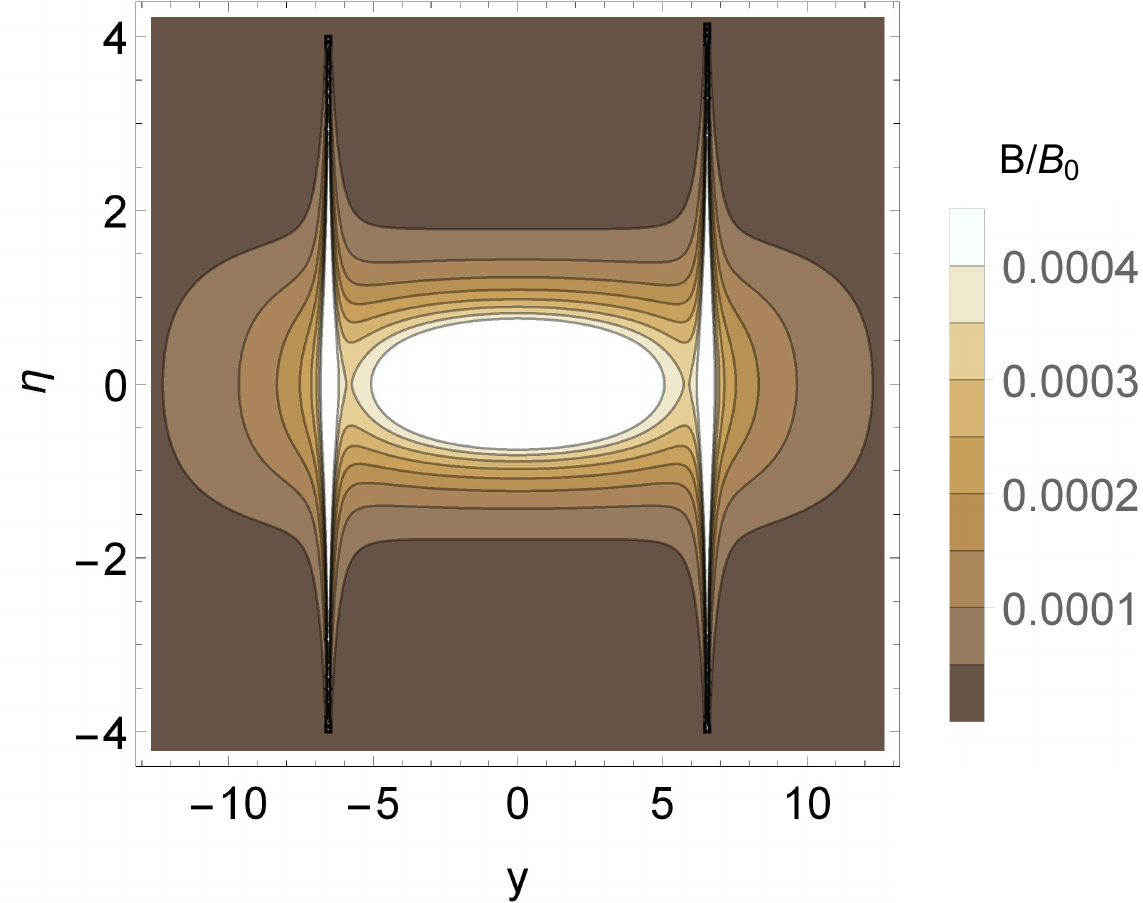}
		\caption{(color online). The magnitude of the magnetic field in $y\!-\!\eta$ plane from current free solution, i.e. Eq.~\eqref{eq:milne-b} using parameters fixed in \eqref{eq:b-params} for $\omega_1=0.5,\,\omega_2=0.05$ at (a) $\tau=0.4\fmc$ and  (b) $\tau=5\fmc$. The value of $B_0$ depends on the choice of parameters, but the overall behavior is independent. The white areas are in the neighborhood of a singularity, and the magnetic field is much larger than other parts.}\label{fig:eByeta}
	\end{figure*}
\begin{figure*}[hbt]\centering	
		\includegraphics[width=7cm]{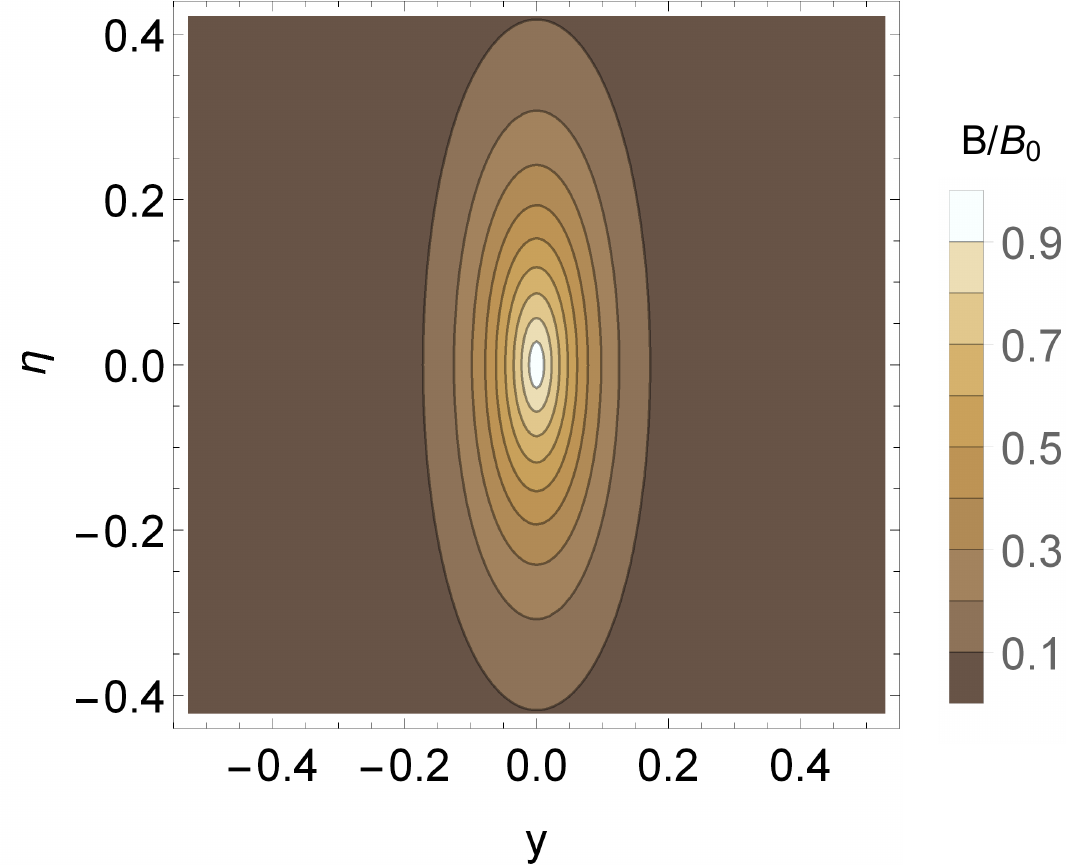}
		\hspace{0.3cm}
		\includegraphics[width=7cm]{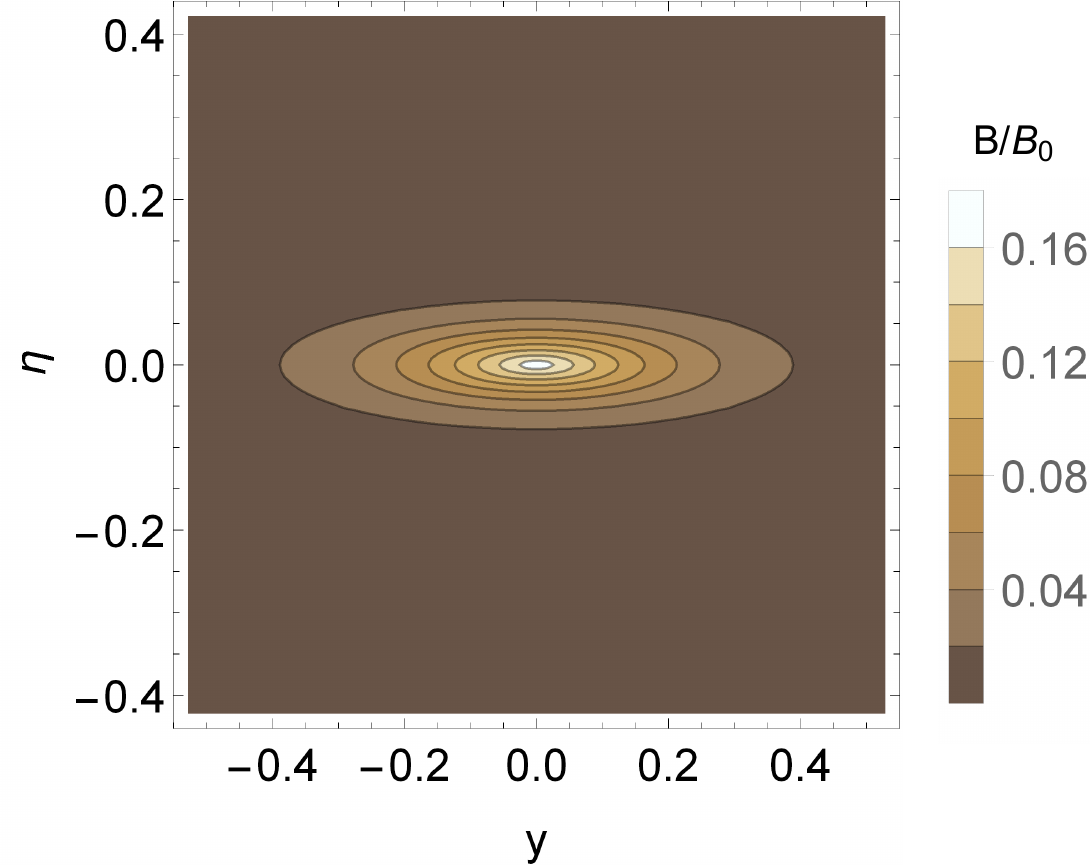}
		\caption{(color online). The magnitude of the regularized magnetic field in $y\!-\!\eta$ plane using parameters fixed in \eqref{eq:b-params} for $\omega_1=0.5,\,\omega_2=0.05$ at (a) $\tau=0.4\fmc$ and  (b) $\tau=5\fmc$.}\label{fig:eByetareg}
	\end{figure*}

	Although the aim of the present work is mainly theoretical, a numerical investigation of the results might be fruitful. To set the stage we choose some benchmark values for $L$ and $T_0$. Following~\cite{Gubser-Conformal}, we first find a relation between the produced entropy and $\Lambda\equiv L T_0$. For a given value of $\Lambda$ the values of $L$ and $T_0$ are arbitrary and depend on the choice of units~\cite{bir}. We then assume the common units in heavy ion physics to fix values of the aforementioned parameters. The physical quantities that we use are~\cite{Inghirami:2019mkc,Gubser-Conformal}
	\begin{equation}\label{eq:physics}
	\epsilon_0=55~\GeV/{\fm^3}=275\fm^{-4}\,,\qquad\tau_0=0.4\fmc\,,\qquad\deriv{S}{\eta}\at[\Big]{\eta=0}=5000\,.
	\end{equation}
	If the freezeout hypersurface is at $\tau=\tau_f$ the entropy per unit rapidity at mid-rapidity reads~\cite{Gubser-Conformal,bir}
	\begin{equation}\label{eq:gubser-formula}
	\deriv{S}{\eta}\at[\Big]{\eta=0}=2\pi\tau_f(2.43)\int_{0}^{{\bf x}_{\perp,f}}\!\!\!{\bf x}_\perp\rd{\xbot} \left(\epsilon^{3/4}u^\tau\right)\at[\Big]{\omega_i=0\,,\eta=0\,,\tau=\tau_f}
	\end{equation}
	At large values of ${\bf x}_\perp$ the integrand decays rapidly 
	\begin{equation*}
	{\bf x}_\perp \left(\epsilon^{3/4}u^\tau\right)\at[\Big]{\omega_i=0\,,\eta=0\,,\tau=\tau_f}\sim \frac{L^4T_0^3}{{\bf x}_\perp^5},\quad\text{for},\quad{\bf x}_\perp\gg L\,,
	\end{equation*}
	while it is linear at small values of ${\bf x}_\perp$
	\begin{equation*}
	{\bf x}_\perp \left(\epsilon^{3/4}u^\tau\right)\at[\Big]{\omega_i=0\,,\eta=0\,,\tau=\tau_f}\sim \frac{L^6T_0^3}{(L^2+\tau^2)^2}{\bf x}_\perp,\quad\text{for},\quad{\bf x}_\perp\ll L\,.
	\end{equation*}
	Plugging the above approximation for the integrand in \eqref{eq:gubser-formula} and using \eqref{eq:physics} gives rise to	
	\begin{equation}\label{eq:Lambda}
	\Lambda\equiv LT_0 = 0.51 \sqrt[3]{\deriv{S}{\eta}\at[\Big]{\eta=0}}\approx 8.68\,. 
	\end{equation}
	If we neglect the rotation in \eqref{eq:milne-epsilon}, then at $\tau_0$ we find
	\begin{equation*}
	\epsilon_0 \approx \frac{90823 L^4}{(L^2+\tau_0^2)^4}\,.
	\end{equation*}
	Plugging numbers from \eqref{eq:physics} and using \eqref{eq:Lambda} gives rise to
	\begin{equation}\label{eq:params}
	L = 4.22\fm,\qquad T_0=2.05~\GeV\,.
	\end{equation}
	The energy density with the above choice of parameters is depicted in Fig.~\ref{fig:epsilonyeta} and Fig.~\ref{fig:epsilonxy}. 
	The current free solution \eqref{eq:milne-b} possesses only one parameter, i.e. $\Sigma_0$. However, due to singularities, it is difficult to choose a value for the aforementioned parameter in a meaningful fashion. One way is to choose the value of $B$ such that at $\left\{\tau=5\fmc\,,x=L\,,y=L\,,\eta=0\right\}$ it satisfies the experimental bound proposed by~\cite{Muller-CME-Bound}
	\begin{equation}\label{eq:muler-bound}
	B\leq 1.35\times\ten{-3}\fm^{-2}\,.
	\end{equation}
	 The other way which we use, is to first regularize the solution with the procedure of Sec.~\ref{sec:regulaization}, fix the parameters and use them in both exact and regularized solutions. To do so, we need to choose the functions $f_\lambda$ and $h_\lambda$ in \eqref{eq:regulrization}. Of course, there are many choices possible for the aforementioned functions. One appropriate choice could be 
	 \begin{equation}\label{eq:reg-choice}
	 f_\lambda(\theta')=\frac{2}{1+\exp\Big[\cos\theta'/\lambda\Big]}\,,\qquad h_\lambda(\theta')=\frac{2}{1+\exp\Big[\sin\theta'/\lambda\Big]}\,.
	 \end{equation}
	  We adopt $\omega_1=0.5$ and $\omega_2=0.05$ from~\cite{bir} and choose $\lambda=\omega_1\omega_2$. The value of $Q_1$ must be chosen in a way such that $\Sigma_0\ll LT_0$ is satisfied. To fix $\Sigma_0$ and $Q_2$ we then assume that the $S^3\times\mathbb{R}$ limits, i.e. \eqref{eq:b-limits} for $B$ at the pole and equator are respectively $\ten{-2}m_\pi^2/e$ and $1.35\times\ten{-3}\fm^{-2}$. With these considerations in mind, the following choice of parameters are suitable
	 \begin{equation}\label{eq:b-params}
	 \Big\{Q_1\,,Q_2\,,\Sigma_0\Big\}=\Big\{100\,,-0.60\,,0.016\Big\}\,.
	 \end{equation} 
	 Although the resulting regularized solution has finite limits at $y=0$ and $y=L$, obtaining exact values for $B$ and $\delta\rho_e$ at the center of the transverse plane is not numerically possible, and therefore we assume a very small value of ${\xbot}_0=\ten{-4}\fm$ to fix a reference point for both $B$ and $\delta\rho_e$. The references values for these quantities are therefore defined as
	 \begin{eqnarray}\label{eq:ref-points}
	 \delta\rho_{e,0}&=&\delta\rho_{e}\left(\tau=2\tau_0,\,x={\xbot}_0,\,y=L,\,\eta=0\right)\sim  3\times\ten{-5}\fm^{-3},\,\nonumber\\ eB_0&=&eB\left(\tau=\tau_0,\,x={\xbot}_0,\,y={\xbot}_0,\,\eta=0\right)\sim 4\times\ten{-2}~m_\pi^2\,.
	 \end{eqnarray}
	 Now we are in a position to represent some figures for the solution. In Fig.~\ref{fig:eByeta}, the magnetic field is depicted in the $y\!-\!\eta$ plane at two different times. In both graphs, there are regions around $y=0$ and $y=L$ in which the magnetic field is very large. As time goes on the shape of the singular regions is modified: It is compressed in the $\eta$ direction and extended in the $y$-direction. Although it is not clear in these pictures, the magnetic field decays with $\tau$, $y$, and $\eta$. In Fig.~\ref{fig:eByetareg}, the same plane in the same instances of time is drawn for the regularized solution. Although the shape of the magnetic field profile becomes regular, the compression and extension are similar to the exact solution of Fig.~\ref{fig:eByeta}. 
\par
\begin{figure*}[hbt]\centering	
		\includegraphics[width=7cm]{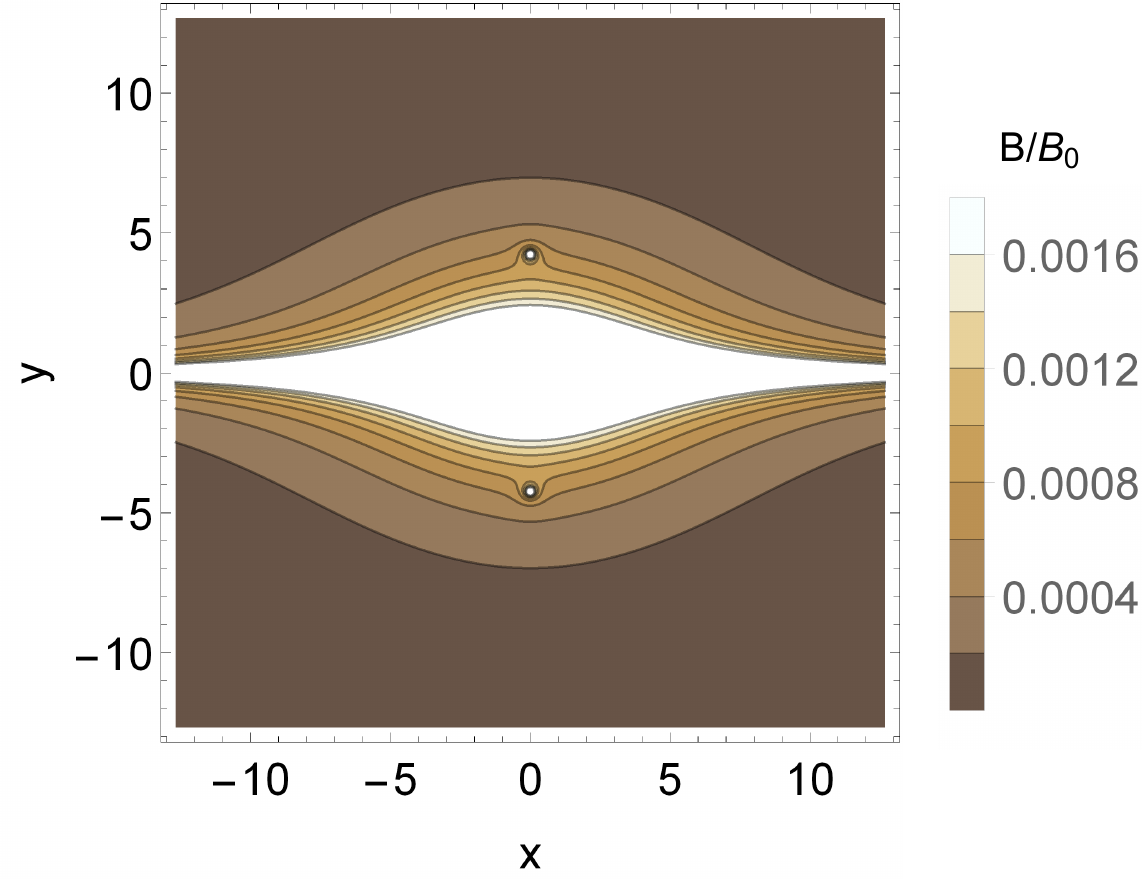}
		\hspace{0.3cm}
		\includegraphics[width=7cm]{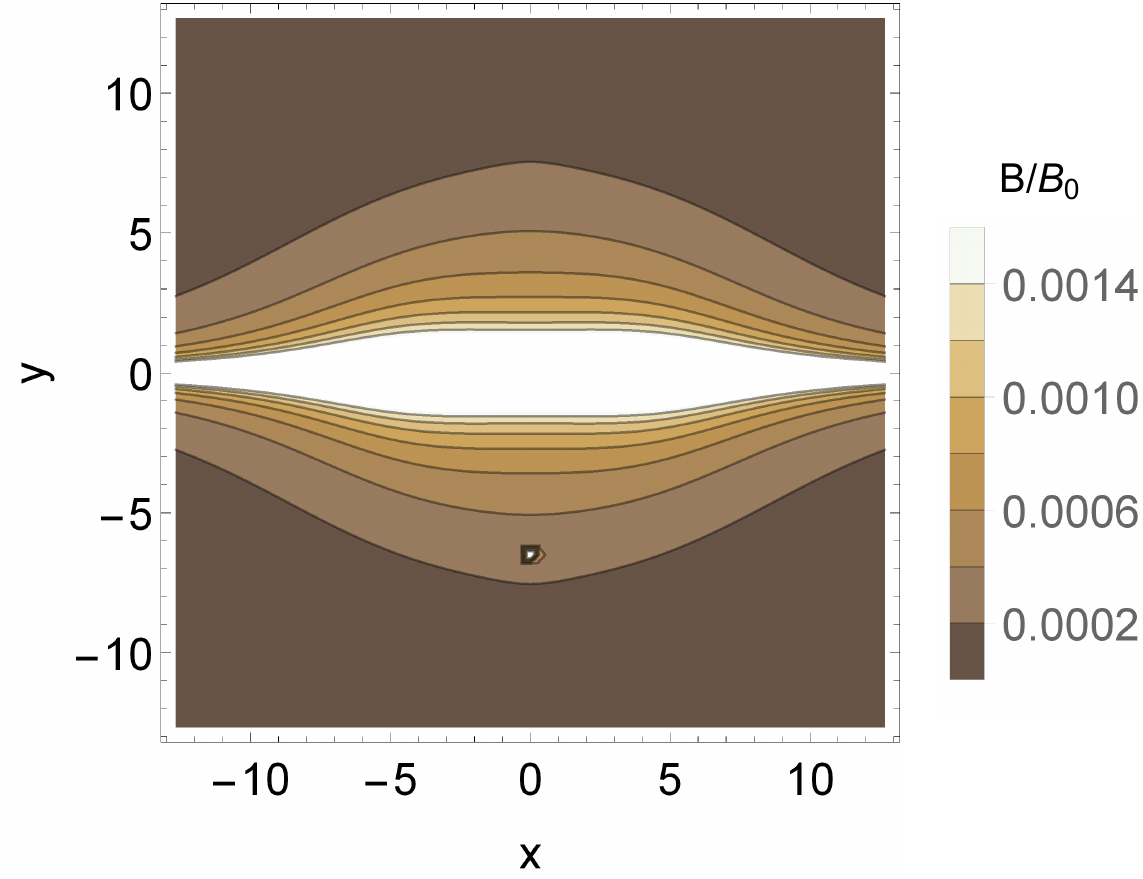}
		\caption{(color online). The magnitude of the magnetic field in the transverse plane from current free solution, i.e. Eq.~\eqref{eq:milne-b} using parameters fixed in \eqref{eq:b-params} for $\omega_1=0.5,\,\omega_2=0.05$ at (a) $\tau=0.4\fmc$, and  (b) $\tau=5\fmc$. The height of the white areas is decreased with time.}\label{fig:eBxy}
	\end{figure*}
\begin{figure*}[hbt]\centering	
		\includegraphics[width=7cm]{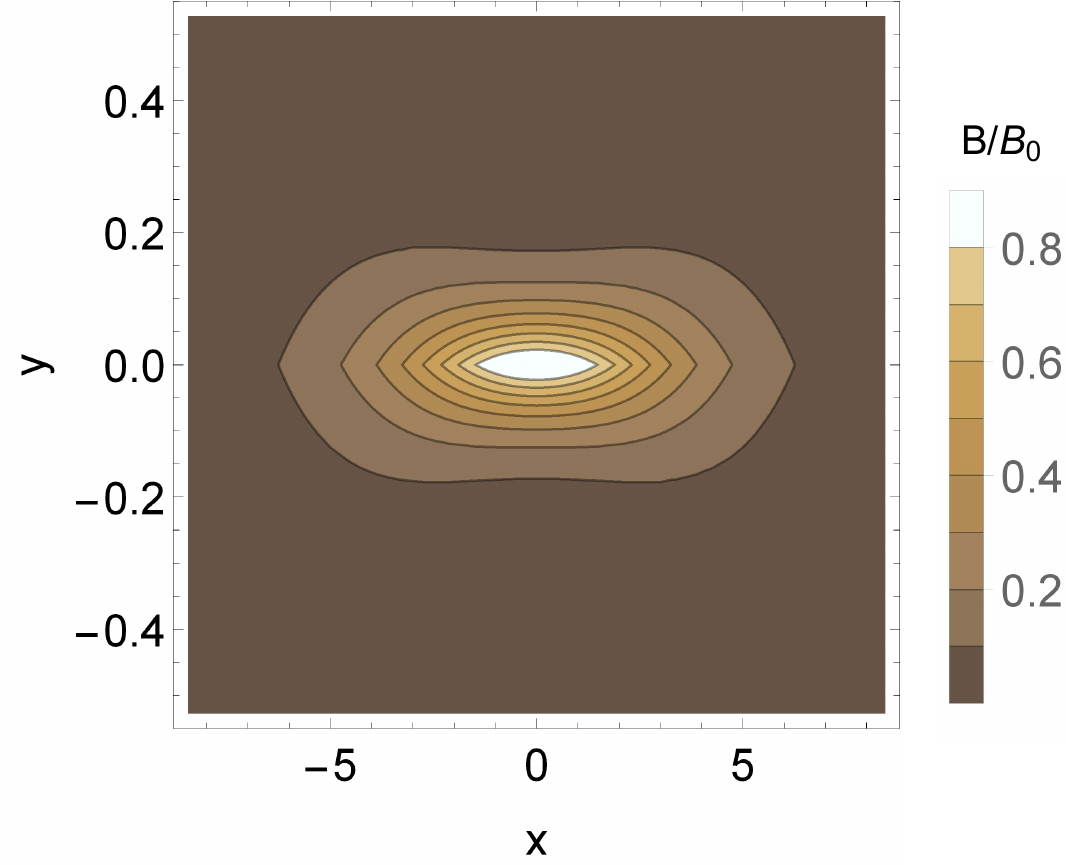}
		\hspace{0.3cm}
		\includegraphics[width=7cm]{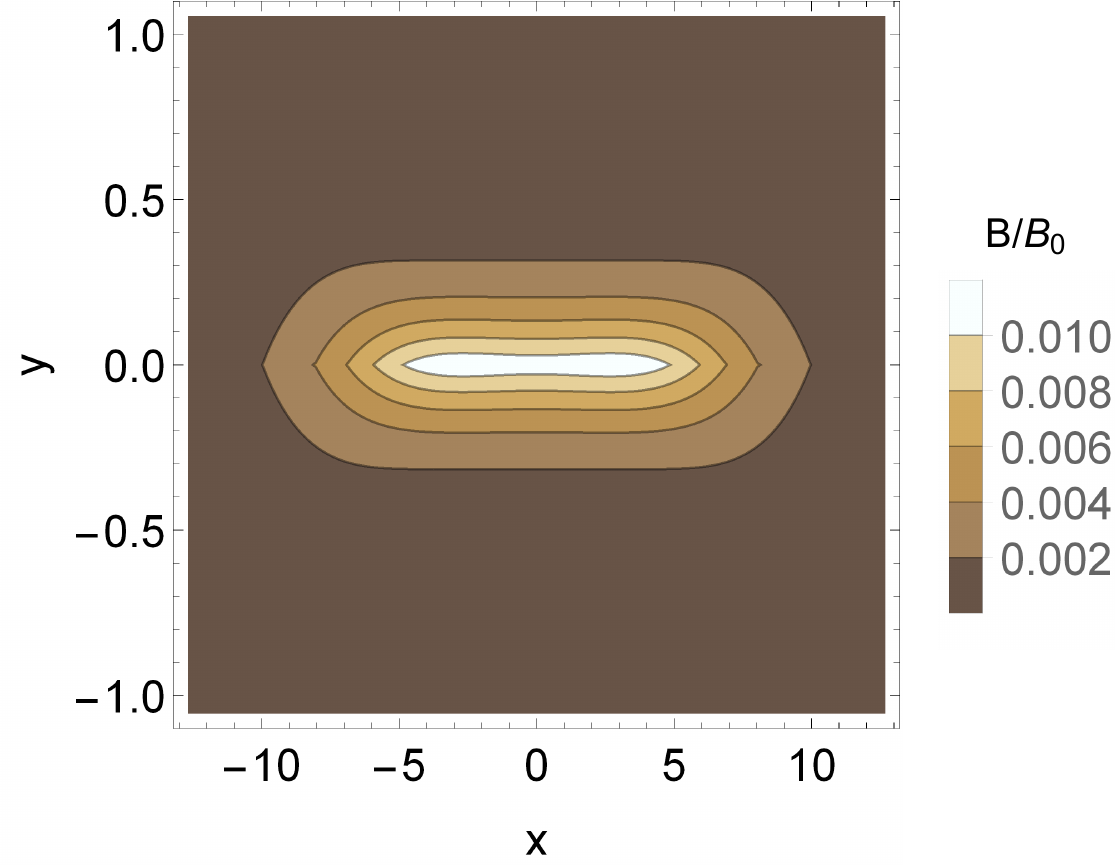}
		\caption{(color online). The magnitude of the regularized magnetic field in the transverse plane using parameters fixed in \eqref{eq:b-params} for $\omega_1=0.5,\,\omega_2=0.05$ at (a) $\tau=0.4\fmc$, and  (b) $\tau=5\fmc$.  }\label{fig:eBxyreg}
	\end{figure*}
\begin{figure*}[hbt]\centering	
		\includegraphics[width=7cm]{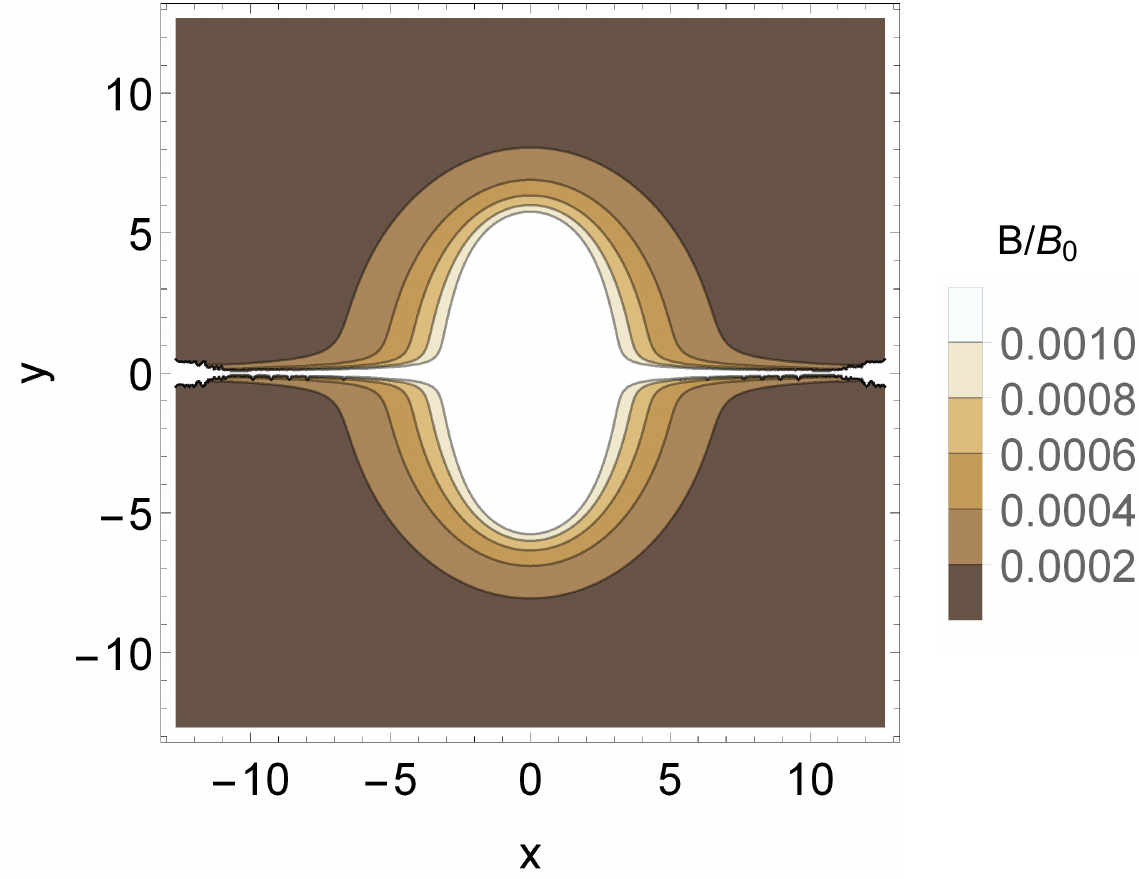}
		\hspace{0.3cm}
		\includegraphics[width=7cm]{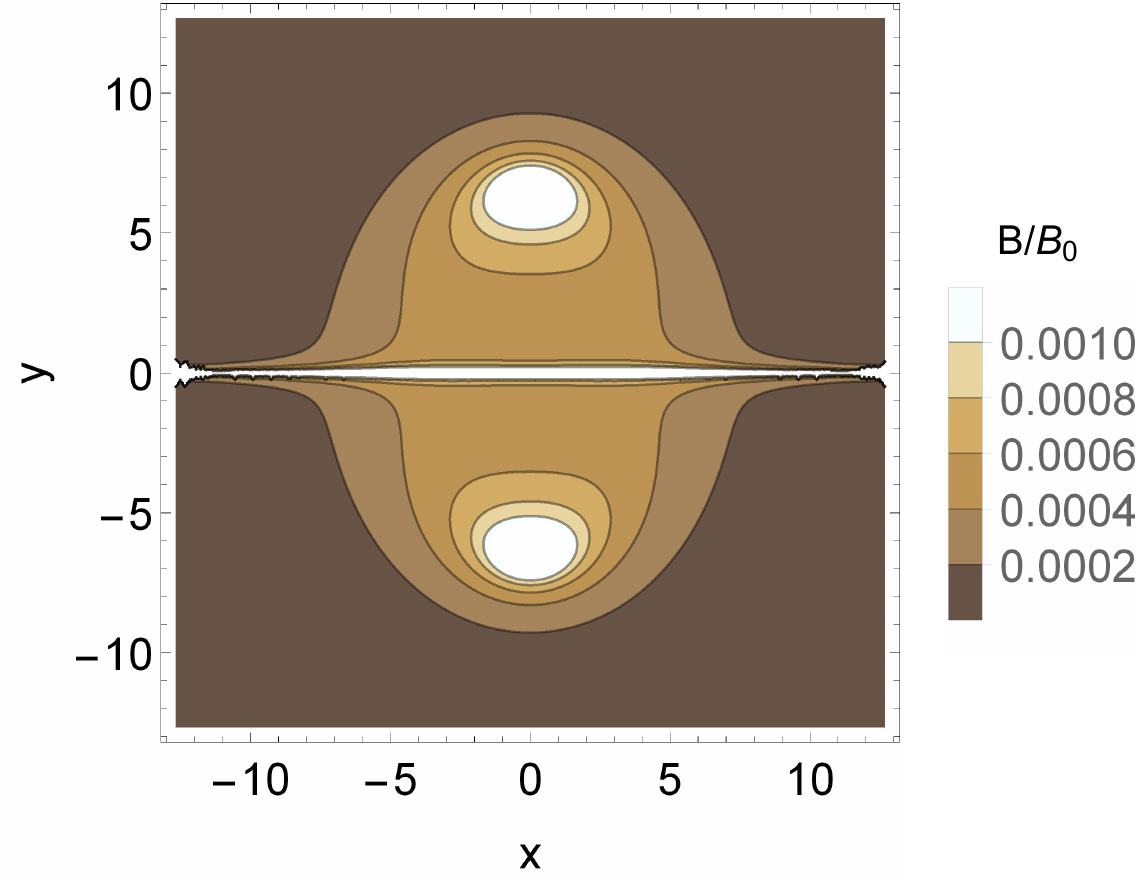}
		\caption{(color online). The magnitude of the magnetic field in the transverse plane from current free solution, i.e. Eq.~\eqref{eq:milne-b} using parameters fixed in \eqref{eq:b-params} for $\omega_1=0.05,\,\omega_2=0.5$ at (a) $\tau=0.4\fmc$, and  (b) $\tau=5\fmc$. The height of the white areas is decreased with time.}\label{fig:eBxy-ex}
	\end{figure*}
\begin{figure*}[hbt]\centering	
		\includegraphics[width=5.1cm]{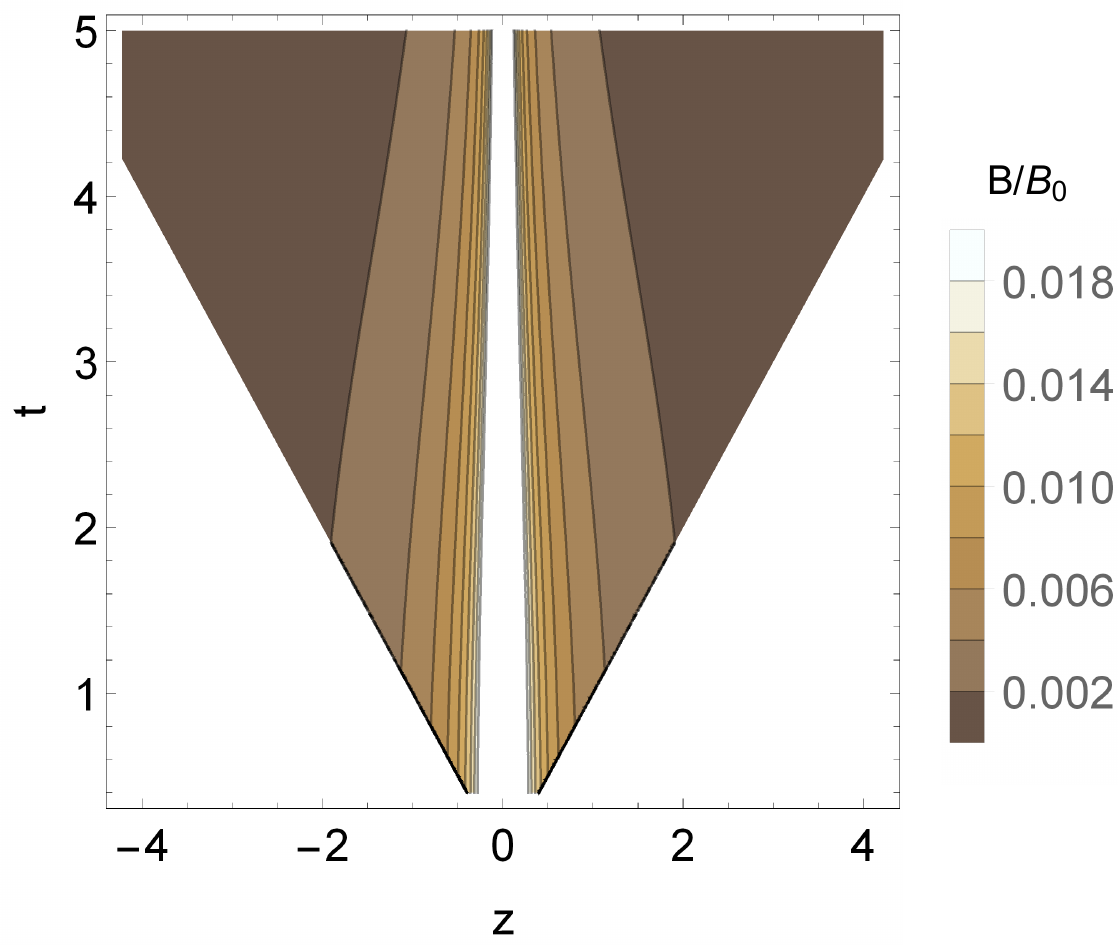}
		\hspace{0.1cm}
		\includegraphics[width=5.1cm]{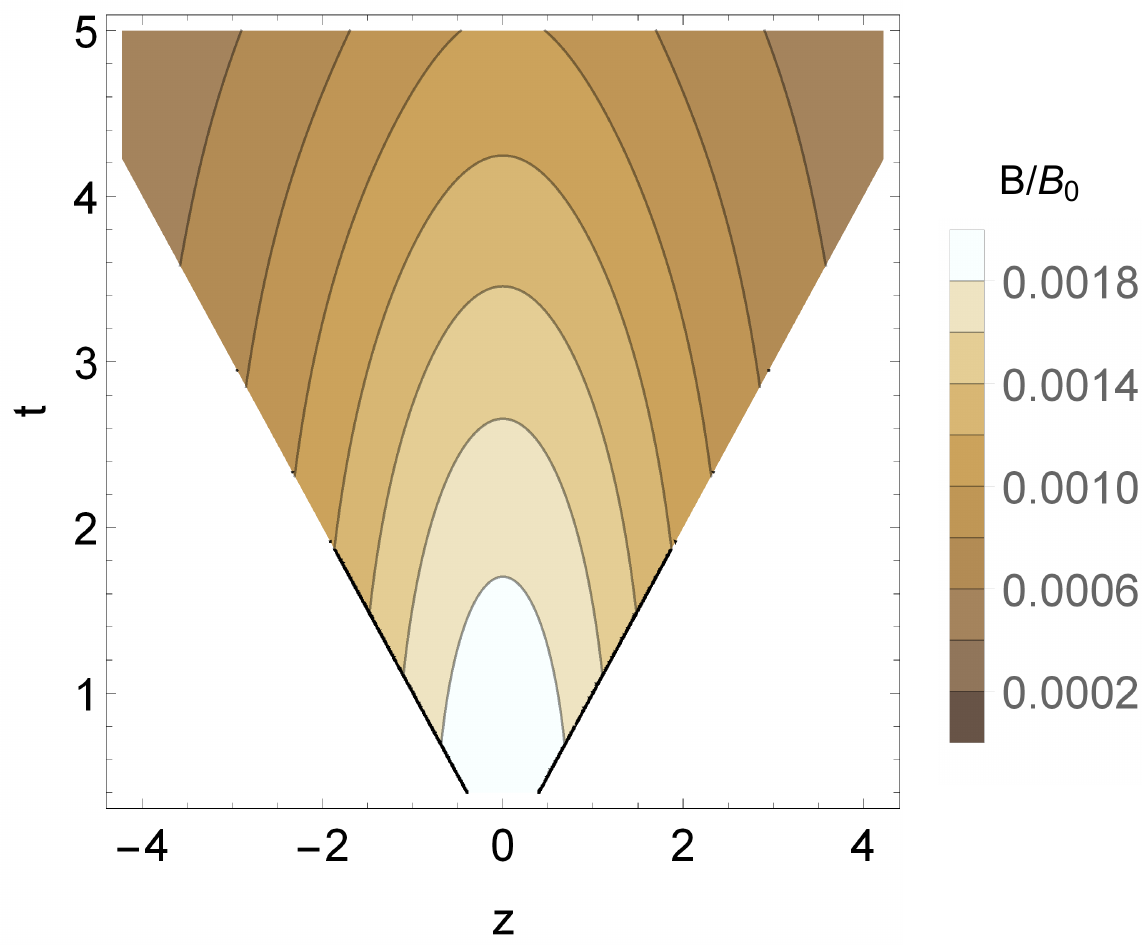}
		\hspace{0.1cm}
		\includegraphics[width=5.1cm]{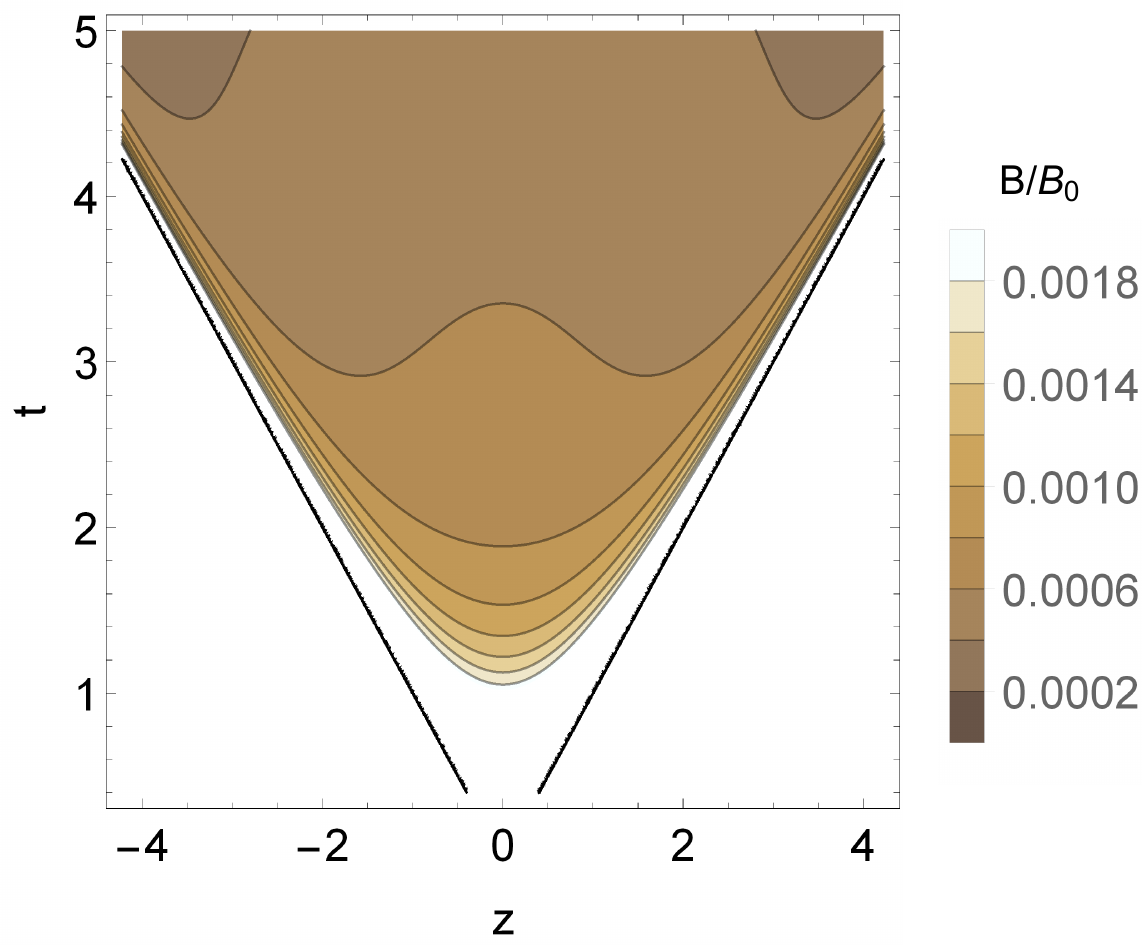}
		\caption{(color online). The magnitude of the magnetic field in the longitudinal ($t\!-\!z$) plane from current free solution, i.e. Eq.~\eqref{eq:milne-b} using parameters fixed in \eqref{eq:b-params} for $\omega_1=0.5,\,\omega_2=0.05$ at (a) $y=0$,  (b) $y=L/2$, and (c) $y=L$. At $y=L$ the earliest iso-magnetic lines resemble the Bjorken iMHD~\cite{RPRR-Bjorken-iMHD,rotating-trans-mhd}. }\label{fig:eBtz}
	\end{figure*}
\begin{figure*}[hbt]\centering	
		\includegraphics[width=7cm]{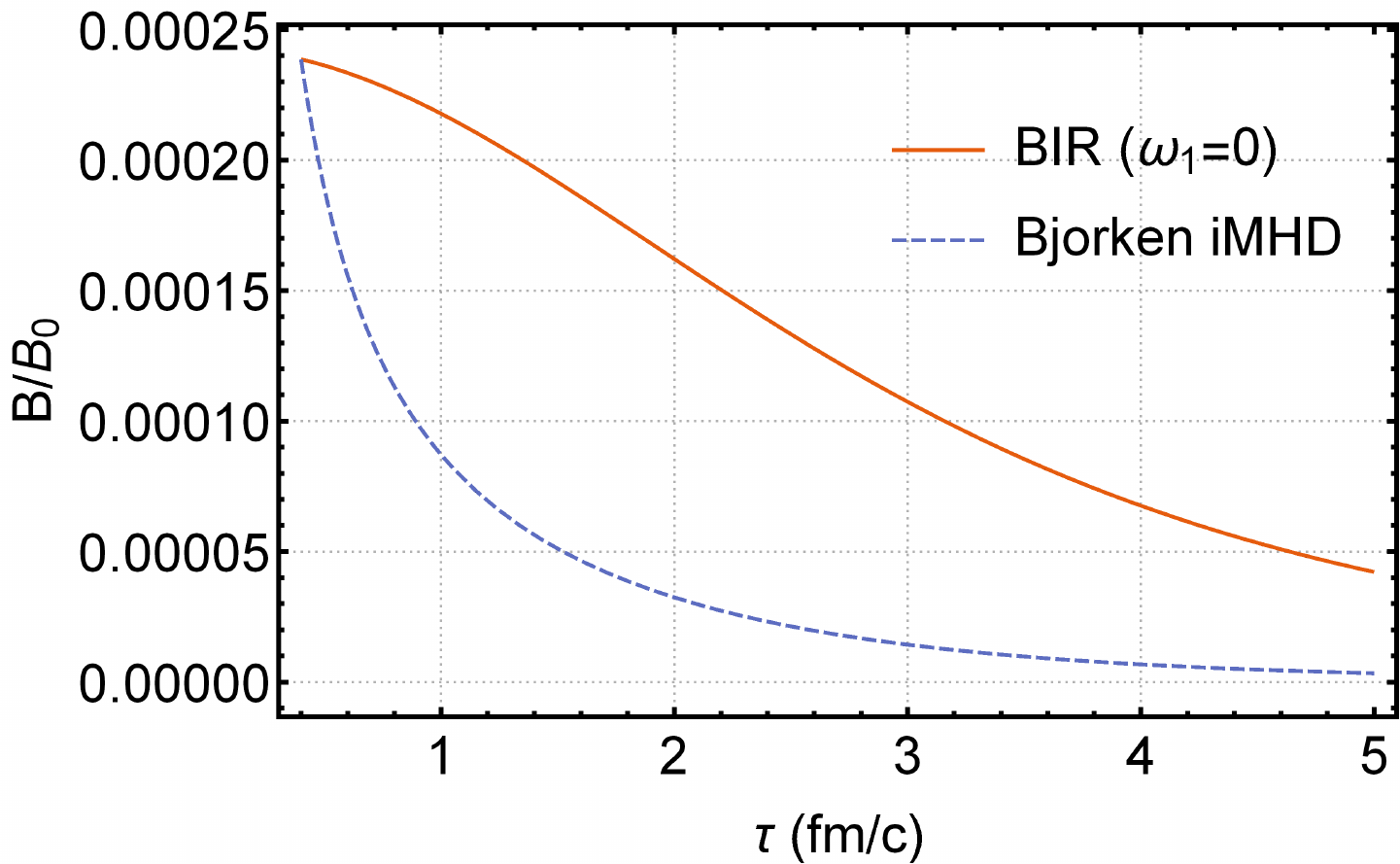}
		\hspace{0.3cm}		
		\includegraphics[width=7cm]{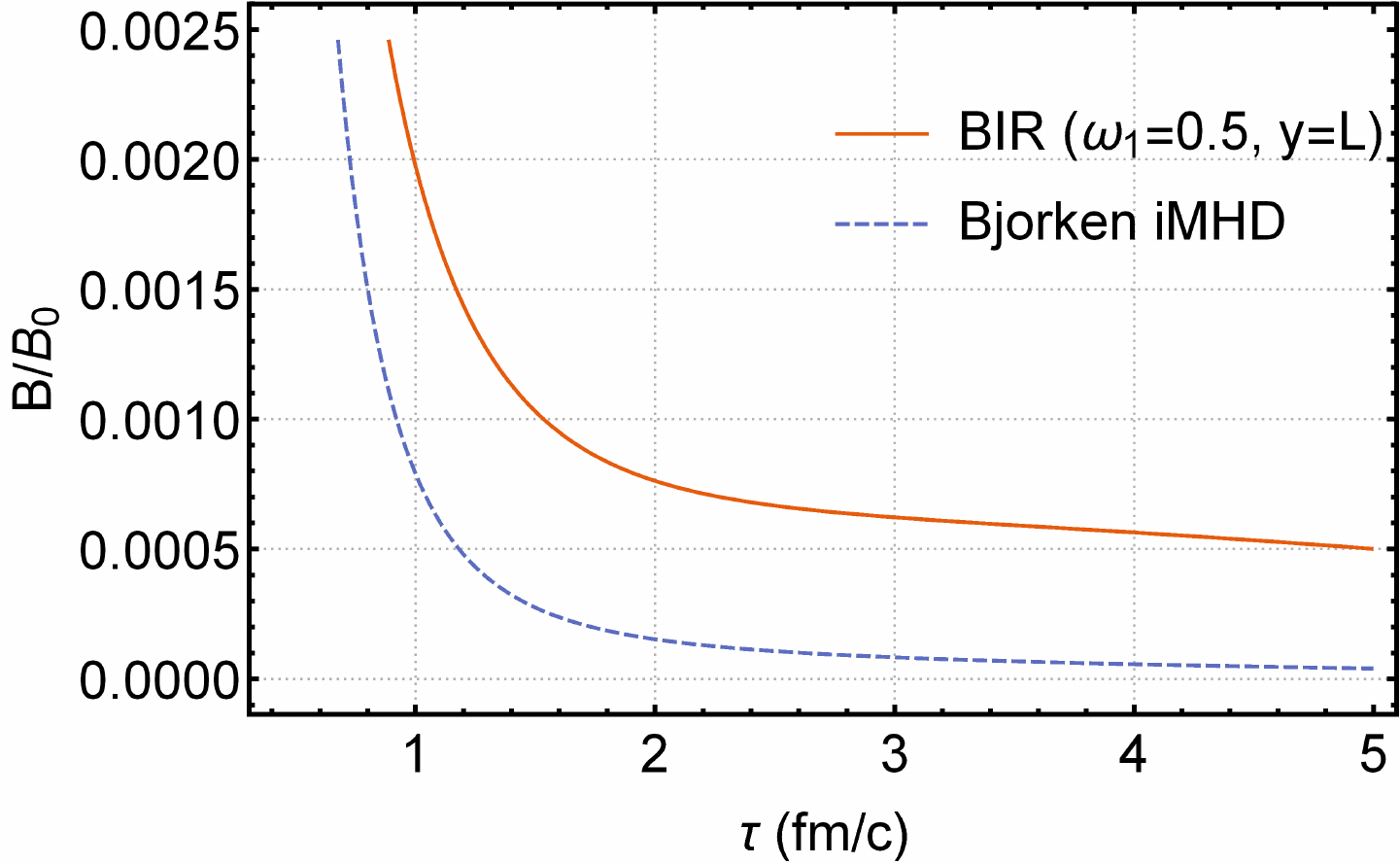}
		\caption{(color online). A comparison of the $\tau$-evolution of the current free solution of Eq.~\eqref{eq:milne-b} with Bjroken iMHD, i.e $B=B_0\frac{\tau_0}{\tau}$. The solid line in the left picture shows the solution with $\left\{\omega_1=0,\,\omega_2=0.05,\,y=0,\,x=0,\,\eta=0\right\}$  while in the right picture it represents the  solution with $\left\{\omega_1=0.5,\,\omega_2=0.05,\,y=L,\,x=0,\,\eta=0\right\}$ is slightly weaker.}\label{fig:eBtau}
	\end{figure*}
\begin{figure*}[hbt]\centering	
		\includegraphics[height=5cm]{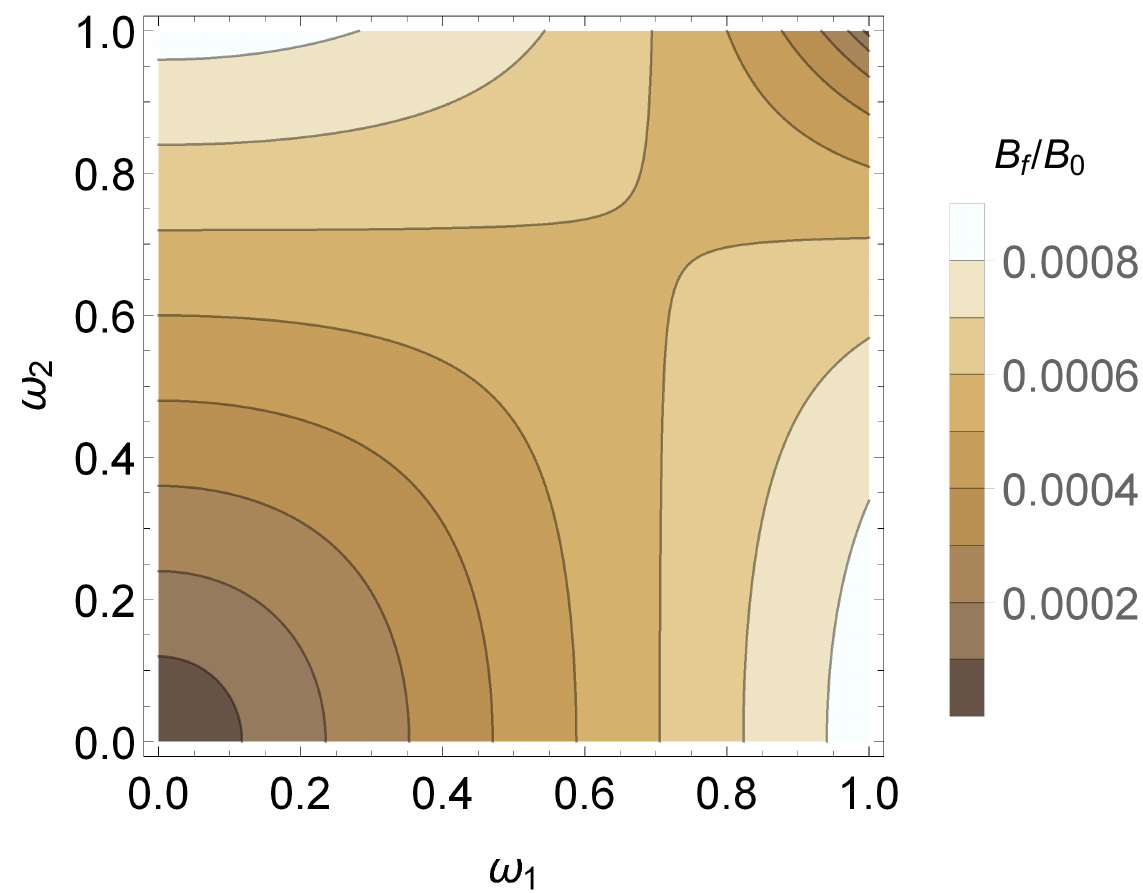}	
		\caption{(color online). (a) Change in $B(\tau=5\fmc,\,x=y=L,\,\eta=0)$ with respect to $\omega_1$ and $\omega_2$. The maximum happens at $\omega_1=\omega_2$ line. (b) Temporal evolution of the regularized magnetic field for different values of $\omega_2$. Other parameters are $\Big\{x=y={\xbot}_0,\,\eta=0,\,\omega_1=0.5\Big\}$}\label{fig:eBomega}
	\end{figure*}

	Fig.~\ref{fig:eBxy} represents the magnetic field of the current free solution in the $x\!-\!y$ plane in two instances of time. In contrast to the $y\!-\!\eta$ plane, the shape of the magnetic field profile is not significantly modified by time. This is also true if we consider the regularized magnetic field as in Fig.~\ref{fig:eBxyreg}. In Fig.~\ref{fig:eBxy-ex},  the values of $\omega_1$ and $\omega_2$ are exchanged and the magnetic field's profile is rotated by $\pi/2$. To represent the temporal evolution of the magnetic field we depict its profile in the $t\!-\!z$ plane in Fig.~\ref{fig:eBtz} as well as a comparison with Bjorken iMHD in Fig.~\ref{fig:eBtau}. The crucial point in Fig.~\ref{fig:eBtz} is that in $y=L$ the early iso-magnetic lines resemble the Bjorken iMHD ones. However, as Fig.~\ref{fig:eBtau} suggests the magnetic field in BIR flow decays much slower than the Bjorkenian case. One may infer that the rotation enhances the magnetic field in the earlier case. The enhancement is more clearly illustrated in Fig.~\ref{fig:eBomega} in which the magnetic field at $\Big\{x=y={\xbot}_0,\,\eta=0,\,\omega_1=0.5\Big\}$ is represented with respect to the angular parameters. For a fixed value of $\omega_1$($\omega_2$) the magnetic field is enhanced by increasing $\omega_2$($\omega_1$) and the maximum is on the $\omega_1=\omega_2$ line.	
	\begin{figure*}[hbt]\centering	
		\includegraphics[height=5cm]{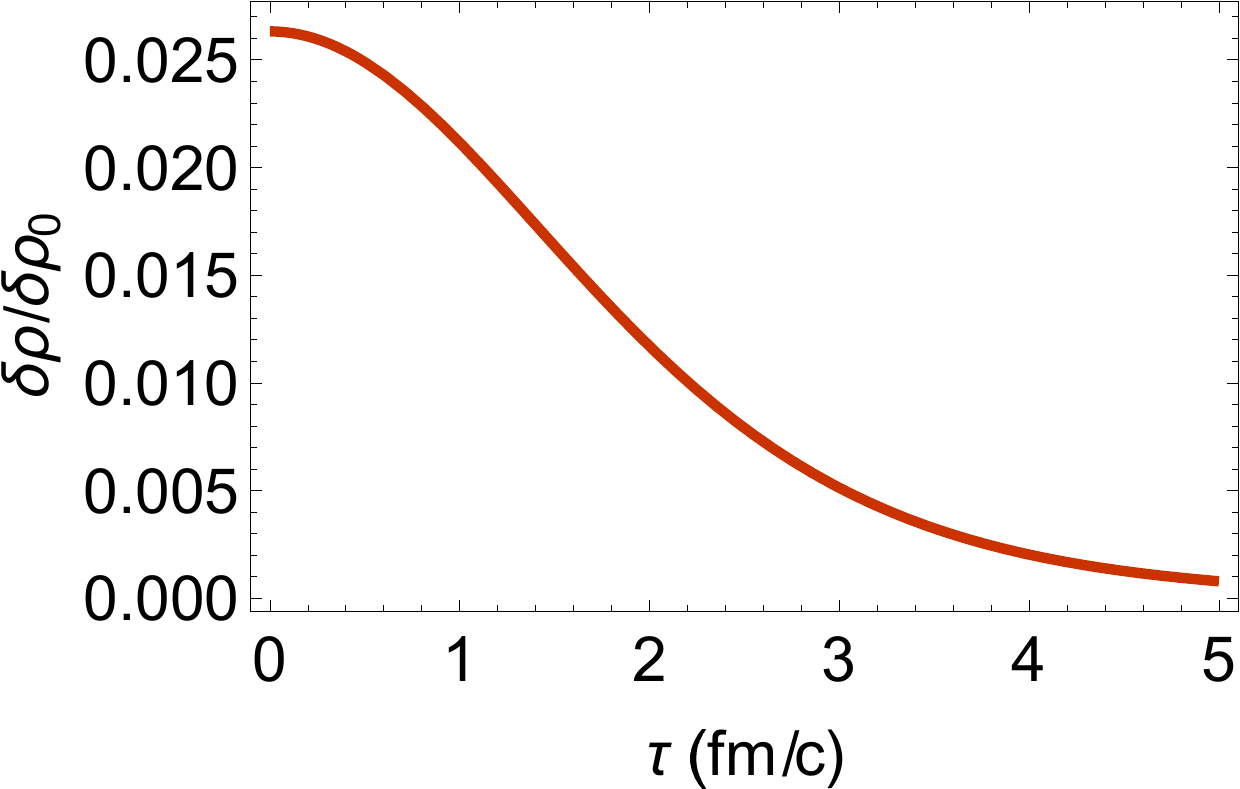}	
		\hspace{0.3cm}		
		\includegraphics[height=5cm]{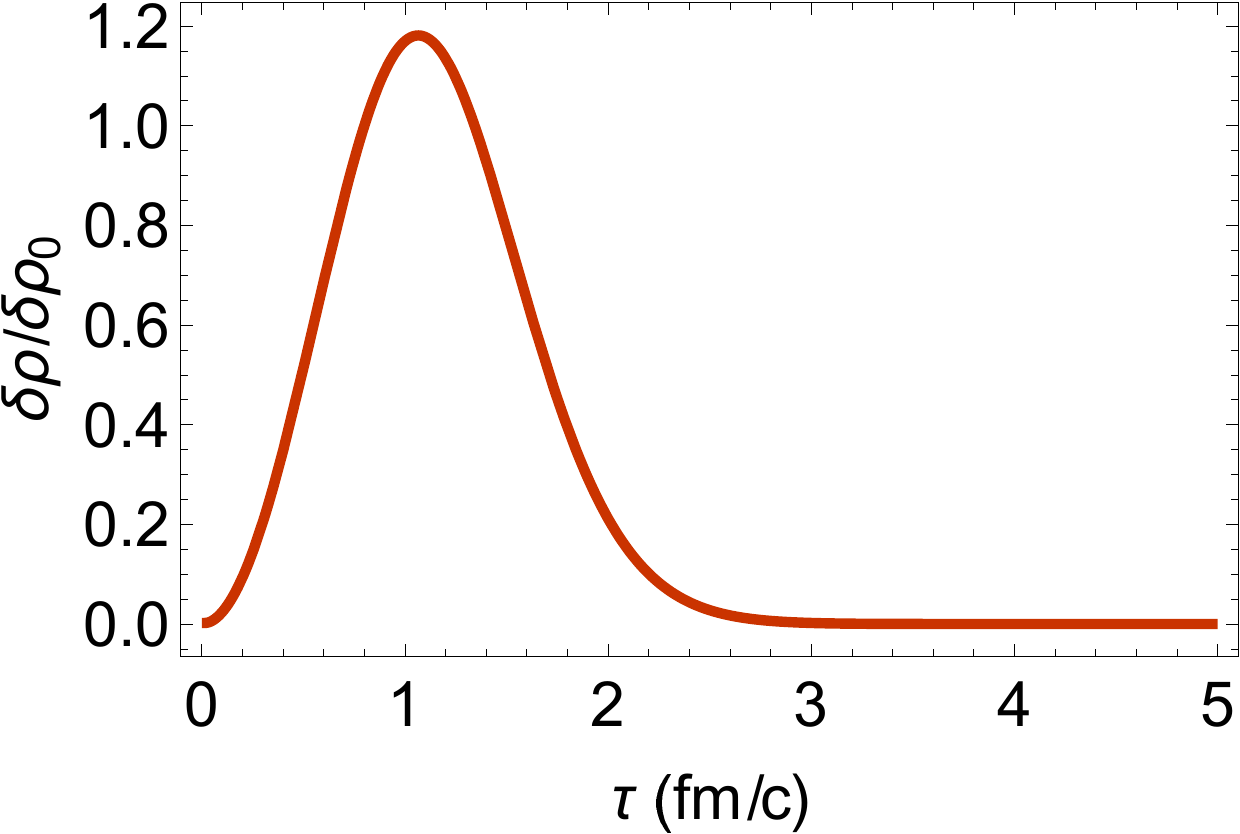}
		\caption{(color online). Temporal evolution of $\delta\rho_e$ with respect to $\delta\rho_{e,0}$ defined in \eqref{eq:ref-points} at (a) $y={\xbot}_0$, and (b) $y=L$. Other parameters are $\Big\{x={\xbot}_0,\,\eta=0,\,\omega_1=0.5,\,\omega_2=0.05\Big\}$}\label{fig:deltarhotau}
	\end{figure*}
	\begin{figure*}[hbt]\centering	
		\includegraphics[width=5cm]{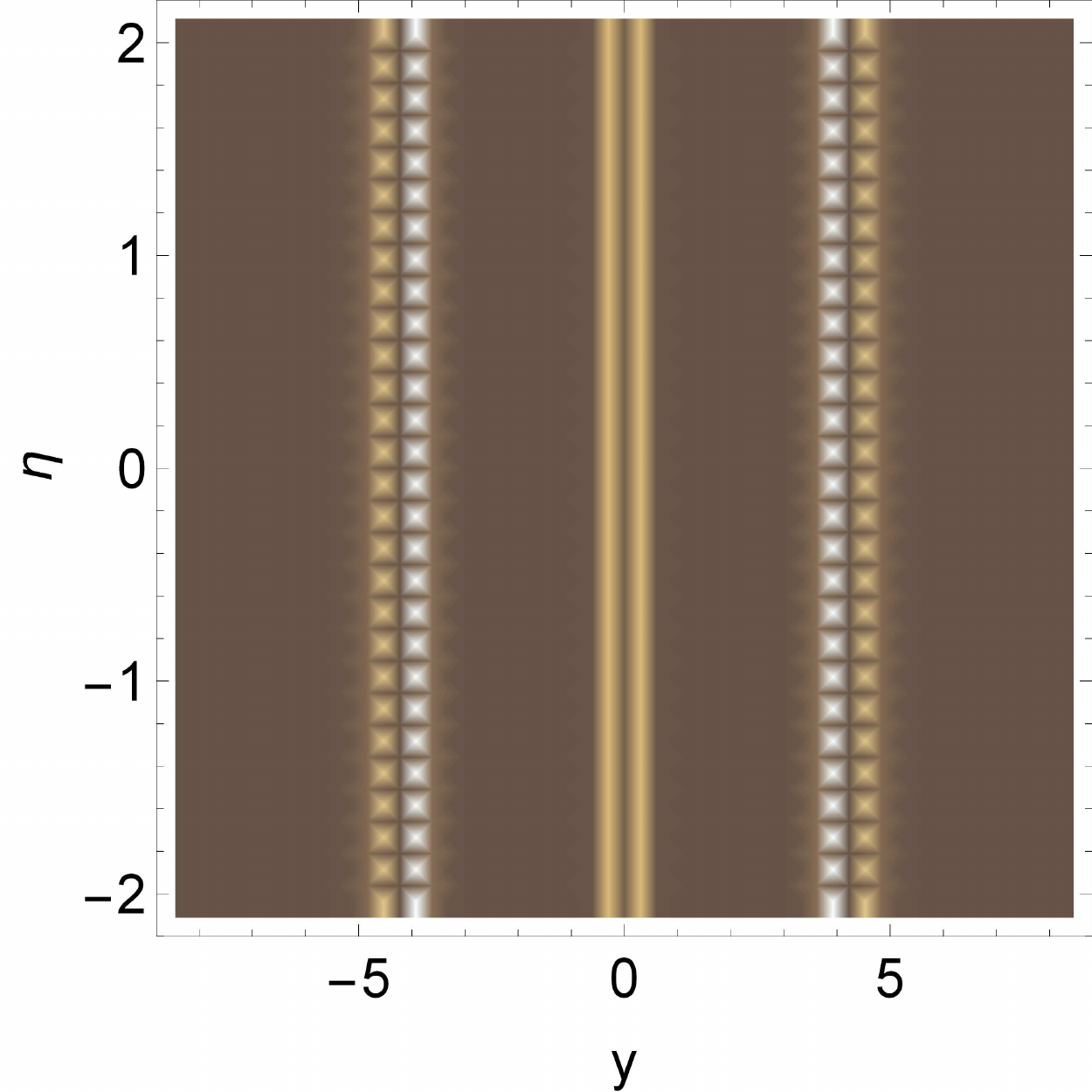}	
		\hspace{0.1cm}		
		\includegraphics[width=5cm]{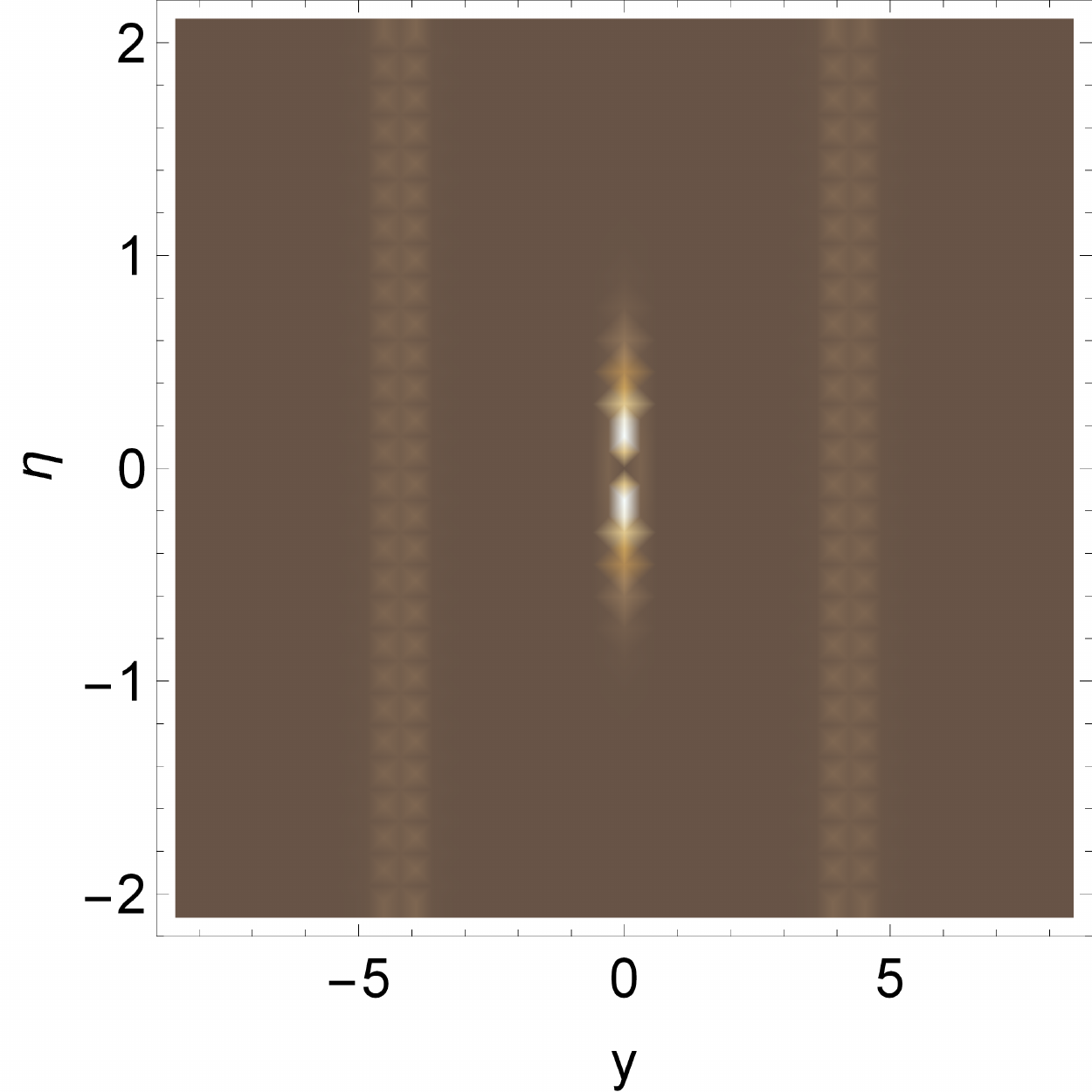}
		\hspace{0.1cm}
		\includegraphics[width=5cm]{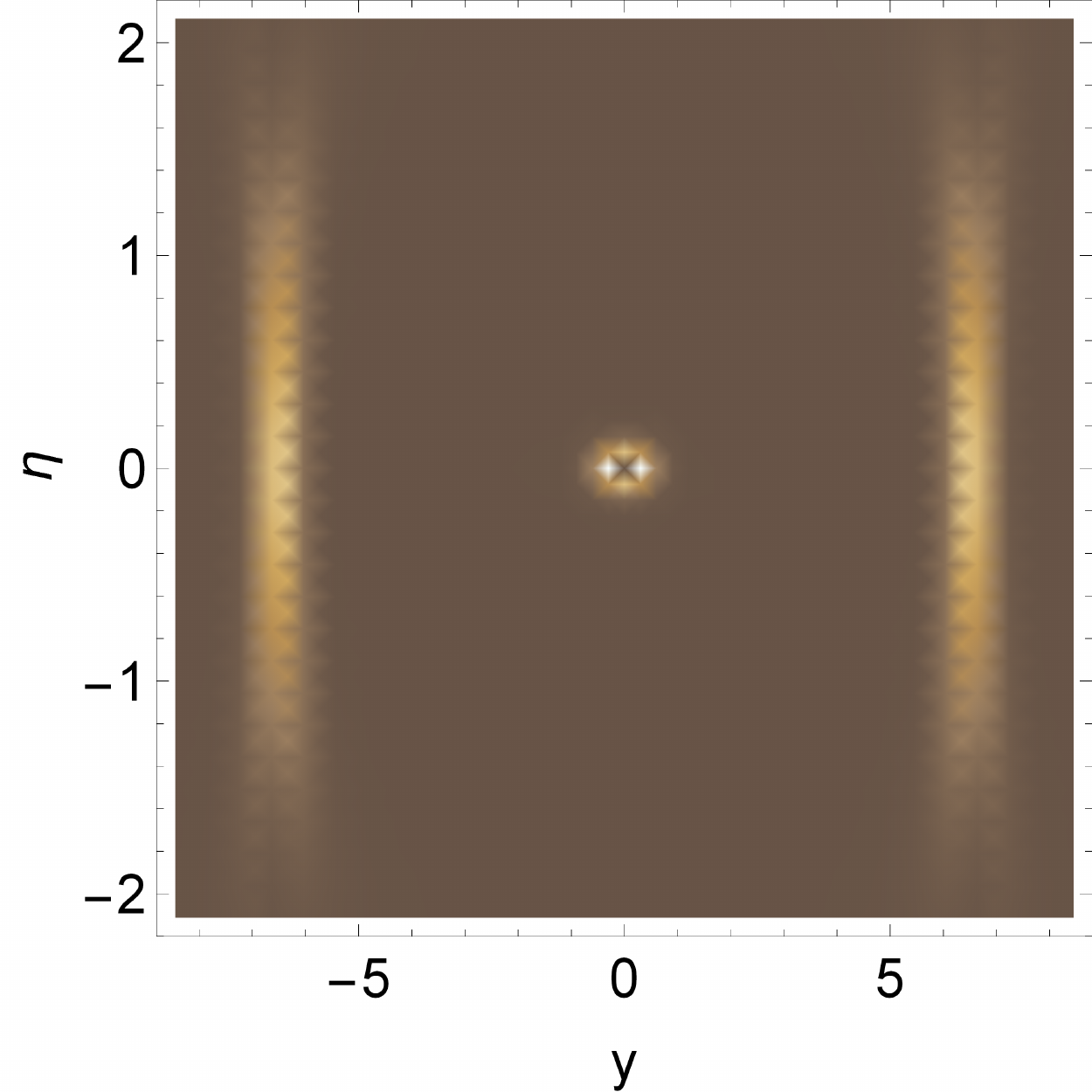}
		\caption{(color online). $\delta\rho$ in the $y\!-\!\eta$ plane at (a) $\tau=0$, (b) $\tau=0.4\fmc$, and (c) $\tau=5\fmc$. The value of $\delta\rho$ is multiplied by $\ten{7}$ to make the picture more clear. Other parameters are $\Big\{x={\xbot}_0,\,\eta=0,\,\omega_1=0.5,\,\omega_2=0.05\Big\}$}\label{fig:deltarhoyeta}
	\end{figure*}
	\begin{figure*}[hbt]\centering	
		\includegraphics[width=5cm]{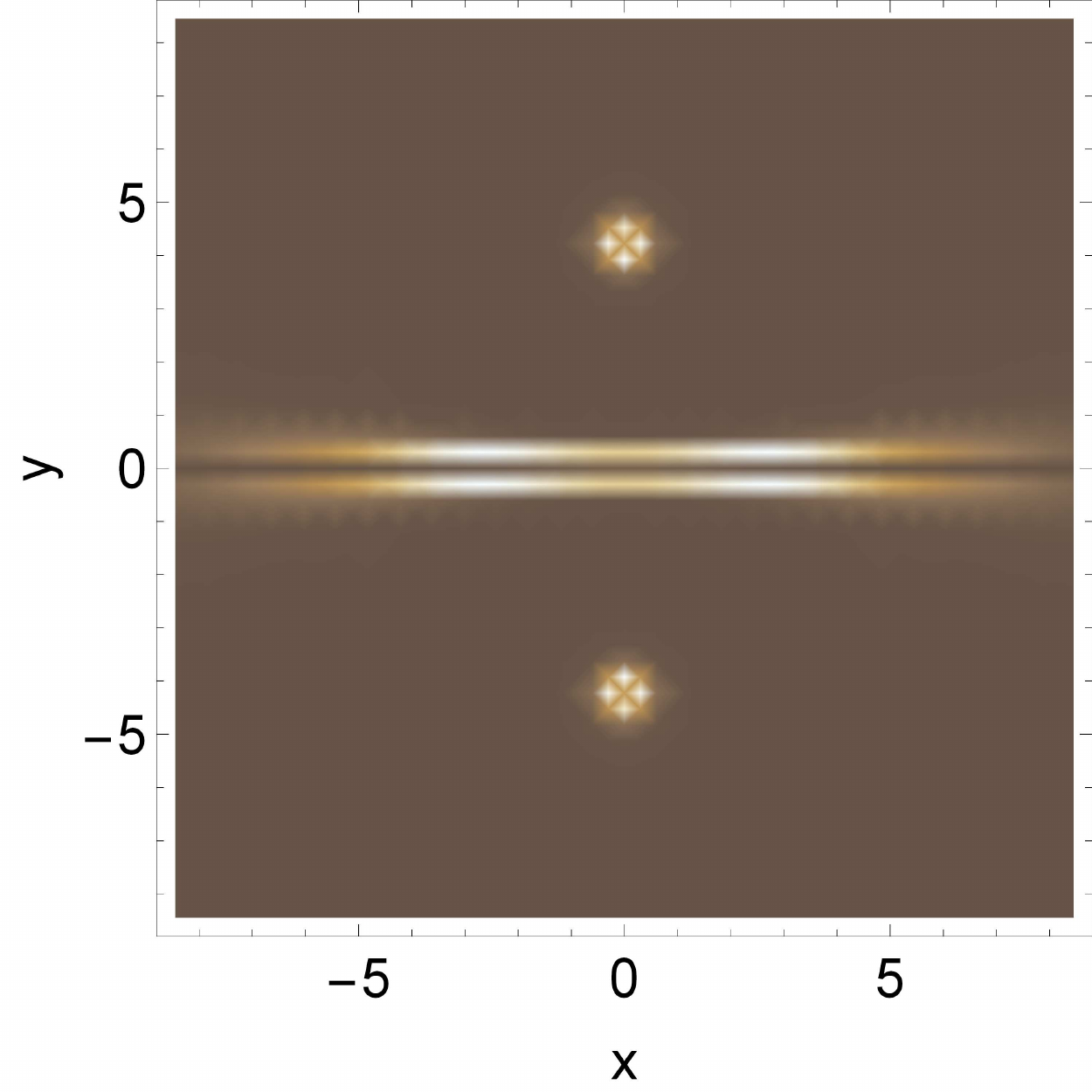}	
		\hspace{0.1cm}		
		\includegraphics[width=5cm]{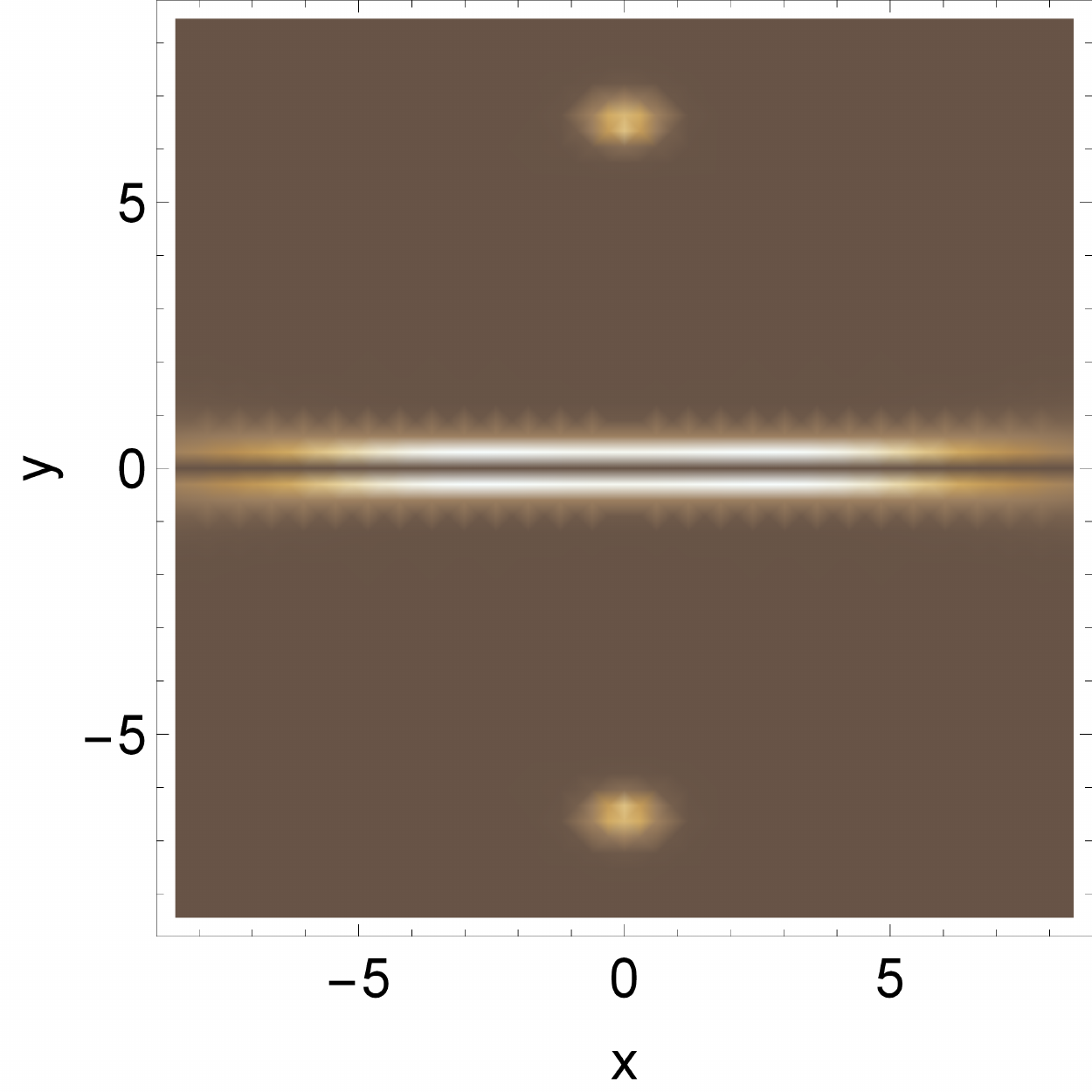}
		\hspace{0.1cm}
		\includegraphics[width=5cm]{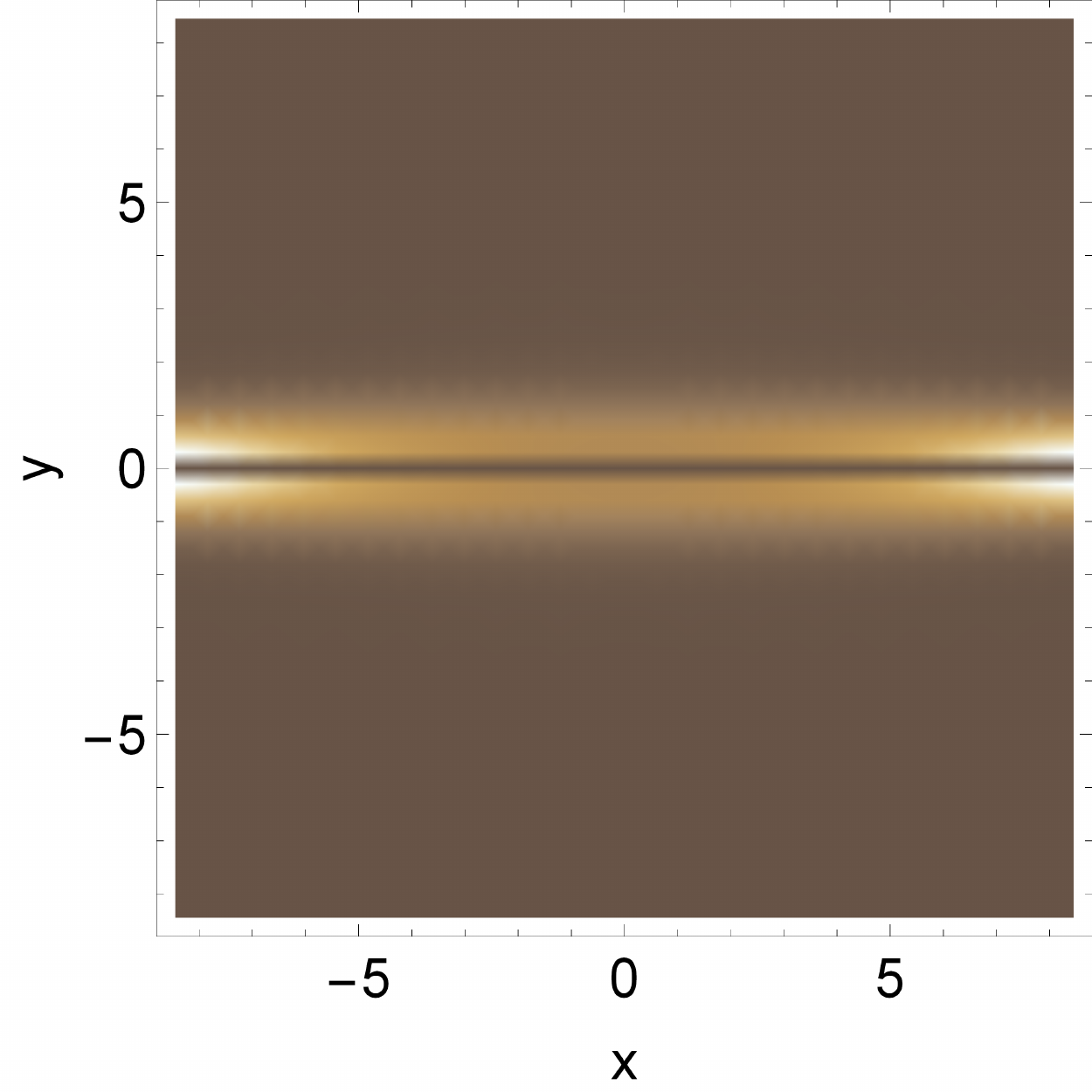}
		\caption{(color online). $\delta\rho$ in the transverse plane at (a) $\tau=0$, (b) $\tau=0.4\fmc$, and (c) $\tau=5\fmc$. The value of $\delta\rho$ is multiplied by $\ten{7}$ to make the picture more clear. Other parameters are $\Big\{x={\xbot}_0,\,\eta=0,\,\omega_1=0.5,\,\omega_2=0.05\Big\}$}\label{fig:deltarhoxy}
	\end{figure*}
\par
As mentioned in Sec.~\ref{sec:regulaization}, the regularization induces a local charge density at the boundaries of $S^3\times\mathbb{R}$ spacetime given in \eqref{eq:s3r-charge}. It worths that we pause and investigate some of the properties of the induced charge density in the flat spacetime. The temporal evolution of the local charge density close to the center of collision, i.e.~$y=0$, and at $y=L$ are depicted in Fig.~\ref{fig:deltarhotau}. In both cases, the charge density relaxes as time goes on. The re-peaking of charge density at $y=L$ is probably a subsequent of our choice for $f_\lambda$ and $h_\lambda$. At both points, the charge is positive which is due to the choice for $Q_2$. One should bear in mind that the value of the reference point, i.e.~$\delta\rho_{e,0}$, and therefore the induced charges are extremely small. Fig.~\ref{fig:deltarhoyeta} is particularly interesting in which the charge at the center is quickly squeezed in the $\eta$-direction to become point-like while the charges at the boundaries are moving outward in the $y$-direction. If Fig.~\ref{fig:deltarhoxy} the charge density is depicted in the transverse plane at mid-rapidity which displays a similar manner. The charges originally located at $y=\pm L$ move outward in the $y$-direction and finally disappear. The central charge fades and spreads through the plane. 
\par
	We should also comment on the direction of the magnetic field in the current solution. Essentially due to the singularities, there are regions of spacetime in which the magnetic field tends to be highly longitudinal, i.e in the z-direction. Since this is probably the artifact of a vanishing current, we deeply doubt that such a high tendency of the magnetic field vector to the longitudinal direction may be likely in any stage of the evolution of the fireball.
	\section{Concluding remarks}\label{sec:conclusion}
	\setcounter{equation}{0}
	\par	
	In the current work, we presented a generalization of the BIR flow to RMHD. For the solution to be manageable, we assumed that the fluid is highly conductive, electrically neutral and the electromagnetic forces are negligible in comparison to the pressure gradient. With such assumptions, we solved the homogeneous Maxwell equation on the boundary of a Kerr-AdS5 black hole by utilization of symmetries. Solving these equations gave rise to the electromagnetic field tensor up to two unknown functions of a single coordinate in the boundary. To determine the aforementioned functions we employed two other assumptions of electrical neutrality and the force-free condition. We showed that these assumptions require the electrical current to be zero. This led to the singularity of the magnetic field at the equator and pole of the boundary, which then translated into the singularities at $y=0$ and $y=L$ in the flat spacetime. By regularizing the results we explicitly showed that these singular behaviors are consequences of the vanishing electrical current. Our solution, albeit being highly theoretical, suggests that the rotation of the fluid may enhance the lifetime of magnetic fields. From a purely theoretical point of view, the present work may contribute to the field of fluid/gravity duality and conformal (magneto)hydrodynamics.
	\par
	In our point of view, further research on the evolution of the electromagnetic fields in the QGP requires reflection concerning the simplifying assumptions that were made in the current work and commonly in the literature. As mentioned in Sec.~\ref{sec:intro}, the iMHD limit may not be proper for the QGP. For a rough estimation let's consider the magnetic Reynolds number, i.e $R_m=Lu\sigma_e$. The typical value of $Lu$ for the QGP is of order $10\fm$, therefore a high $R_m$ regime requires $\sigma_e$ to be larger than $20\MeV$ which is not met by LQCD results~\cite{LQCD-Sigma}. It is also suggested in~\cite{rotating-trans-mhd} that the ratio of electric to magnetic field decays as $e^{-\sigma_e\tau}$, and thus the LQCD results do not support the suppression of the electric field. This may signal that the role of the electric field on an event-by-event basis is highly ignored. It is also assumed that the QGP is locally neutral essentially because it is highly conductive. One may argue that the relaxation time for the charge density is of order $1/\sigma_e$, and it is plausible that the local charge density does not relax in the QGP lifetime. These two considerations may have significant consequences for our understanding of the evolution of the electromagnetic fields and more importantly the search for CME signals. The force-free condition may also become a poor approximation in certain regions of the fluid, and if that is the case a new window for possible modifications of some observables of heavy-ion collisions might be uncovered. In particular, if most of the strength of the magnetic field is lost in the pre-equilibrium stage then there might be a relation between the hydrodynamization time and potential effects of magnetic fields on the early-time observables.
	\par
	We also would like to briefly comment on the BIR solution itself. At the time of writing, this is the only solution found from the fluid/gravity conjecture that is exact and also breaks both boost and rotational invariances assumed by Bjorken and Gubser flows. Also in comparison with the latter flow, the BIR solution comes from a deeper theoretical framework. However, Gubser flow has a particular property that is not present in the BIR solution. It has a single parameter, related to the inverse of fluid transverse size, that when tended to zero Bjorken flow and its symmetries are reproduced. Unfortunately, the BIR solution is not manifestly related to the geometrical properties of the collision and the fluid and cannot reproduce Gubser or Bjorken flow. Looking for a similar solution with the aforementioned properties may be a possible direction of research.
	\acknowledgments
	The author thanks H.~Arfaei, H.~Bantilan, A.~Naseh, L.~Rezzolla, N.~Sadooghi and F.~Taghinavaz for fruitful discussions. He is also particularly grateful to F.~Elahi for reading the manuscript and providing useful comments.
	\bibliographystyle{JHEP}
	\providecommand{\href}[2]{#2}\begingroup\raggedright

\end{document}

%% file: hep-commands.tex
\newcommand{\fm}{\mathop{\rm fm}\nolimits}

\newcommand{\fmc}{\mathop{\rm fm/c}\nolimits}
\newcommand{\ten}[1]{10^{#1}}

\newcommand{\rd}[1]{\mathop{\mathrm{d}#1}}

\newcommand{\deriv}[2]{\frac{{d}#1}{{d}#2}}
\newcommand{\derivv}[1]{\frac{\mathop{{d}}}{\rd#1}}

\newcommand{\pderivv}[1]{\frac{\partial}{\partial#1}}

\newcommand{\inv}[1]{\frac{1}{#1}}

\newcommand{\abs}[1]{\left|#1\right|}

\newcommand{\dform}[1]{\mathbf{#1}}
\newcommand{\cd}[2]{\mathrm{\nabla}_{#1}{#2}}

\newcommand{\lieder}[2]{\mathcal{L}_{#1}{#2}}
\newcommand{\christoffel}[3]{\Gamma^{#1}_{#2#3}}
\newcommand{\at}[2][]{#1|_{#2}}

\newcommand{\dual}[1]{{^\star{#1}}}
\newcommand{\dfdual}[1]{{\star{\dform{#1}}}}
\DeclareRobustCommand{\orderof}{\ensuremath{\mathcal{O}}}

\DeclareMathOperator{\dext}{\mathbf{d}}

\DeclareBoldMathCommand{\xbot}{{\bf x}_\perp}
\newcommand{\tvec}[1]{\boldsymbol{#1}}